\documentclass[12pt,letterpaper,pdftex]{article}
\usepackage{amsthm,amsmath,amssymb,upgreek,bm}
\let\putexOriginalRe\Re
\let\putexOriginalIm\Im
\let\putexOriginalCoth\coth
\let\putexOriginalSquare\square
\let\putexOriginalOverleftrightarrow\overleftrightarrow
\usepackage{putex}
\let\Re\putexOriginalRe
\let\Im\putexOriginalIm
\let\coth\putexOriginalCoth
\let\square\putexOriginalSquare
\let\overleftrightarrow\putexOriginalOverleftrightarrow

\let\Disc\relax

\usepackage{graphicx}
\usepackage{units}
\usepackage{float}
\usepackage[export]{adjustbox}
\usepackage{xcolor}
\usepackage[shortlabels]{enumitem}
\usepackage[mathscr]{euscript}
\usepackage[normalem]{ulem}
\usepackage{tikz}
\usepackage{tikzsymbols}
\usepackage{pgfplots}
\usepackage{ytableau}
\usepackage{slashed}
\usepackage{mathtools}
\usepackage{cite}
\usepackage{hyperref}
\usetikzlibrary{decorations.pathmorphing}
\usetikzlibrary{arrows.meta}
\usepackage{microtype}

\renewcommand{\TitleHook}{\LARGE\bfseries}

\newcommand{\pa}{\partial}
\renewcommand{\hat}{\widehat}
\DeclareMathOperator{\re}{Re}
\DeclareMathOperator{\im}{Im}

\DeclareMathOperator{\Tr}{Tr}
\newcommand{\Disc}{\mathord{\mathrm{Disc}}}

\DeclareMathOperator{\erfc}{erfc}

\newcommand{\la}{\langle}
\newcommand{\ra}{\rangle}

\numberwithin{equation}{section}

\def\bZ {\mathbb{Z}}

\def\be{\begin{equation}}
\def\ee{\end{equation}}
\def\ie{\begin{equation}\begin{aligned}}
\def\fe{\end{aligned}\end{equation}}
\def\ieg{\begin{equation}\begin{gathered}}
\def\feg{\end{gathered}\end{equation}}
 
\newcommand{\cA}{{\mathcal A}}
\newcommand{\cB}{{\mathcal B}}
\newcommand{\cC}{{\mathcal C}}

\newcommand{\cH}{{\mathcal H}}

\newcommand{\cK}{{\mathcal K}}
\newcommand{\cL}{{\mathcal L}}
\newcommand{\cN}{{\mathcal N}}
\newcommand{\cO}{{\mathcal O}}

\newcommand{\cQ}{{\mathcal Q}}
\newcommand{\cR}{{\mathcal R}}

\title{A precision bound for gravitational transseries}
\authors{Ji Hoon Lee}
\institution{eth}{\unexpanded{\shortstack{Institut f\"{u}r Theoretische Physik, ETH Zurich,\\ CH-8093 Z\"urich, Switzerland}}}
\date{}

\abstract{
We propose that the large $N$ transseries of a holographic observable above the black hole threshold admits a precision bound, a limit on how precisely the transseries can determine the observable at finite $N$, set by the data of its D-brane instanton sectors. In doing so, we develop methods in large $N$ resurgence, based on the work of \'{E}calle, that capture the $e^{-N^2}$ asymptotics of gravitational saddles beyond the $e^{-N}$ instantons. Under an assumption on the growth of the D-brane instanton sums that we expect in the black hole phase, we find an analytic obstruction to the resummation of the large $N$ transseries. This obstruction sets the precision bound. As our main example and consistency check, we study the $\frac{1}{16}$-BPS index of $\cN=4$ super Yang-Mills and its black hole. We derive relations between the D-brane instanton data on thermal AdS and on the black hole, in the form of a duality that exchanges the numbers of D-brane instantons on the two backgrounds. In the index, these relations reproduce the actions of D3-branes wrapped on the black hole. We conclude with our expectations for generic situations and their relation to the double-cone wormhole.
}

\begin{document}
\hypersetup{pageanchor=false}
\maketitle
\hypersetup{pageanchor=true}

\setcounter{tocdepth}{2}

\tableofcontents

\newpage

\section{Introduction}

Consider an observable $\cO_N$ in a holographic theory of $N \times N$ matrices. This observable $\cO_N$ is defined exactly at finite $N$ via the boundary theory. On the other hand, in the bulk, methods that are presently known give us an expansion of $\cO_N$ at large $N$. How much of the finite $N$ answer $\cO_N$ can the result of a bulk computation actually determine?

To make this question precise, let us first be explicit about what the presently-known methods in the bulk can provide. For concreteness, we take $\cO_N$ to be the thermal partition function $Z_N$ of a theory of $N \times N$ matrices. Its large $N$ transseries $Z_N^{\rm grav}$ in the 't Hooft limit is given by the formal expression,\footnote{We should note that this form of the large $N$ transseries has not been established in general. This expression reflects the types of non-perturbative asymptotics that are presently known and computable in theories of $N \times N$ matrices.}
\be \label{eq: grav}
Z_N \, \sim \, Z_N^{\rm grav} = \sum_{M} e^{-N^2 I_M} \sum_{\alpha_M} e^{-N A_{\alpha_M}} \, \widetilde\Phi_{M,\alpha_M}(1/N).
\ee
In bulk terminology, the notation is as follows. $I_M$ is the on-shell action of a Euclidean gravity saddle $M$, relative to the ``vacuum'' saddle free energy $I_{\rm vac}$ that we normalized as $I_{\rm vac} = 0$; $A_{\alpha_M}$ is the on-shell action of a D-brane instanton $\alpha_M$ on the gravity saddle $M$; $\widetilde\Phi_{M,\alpha_M}(1/N)$ is the large $N$ perturbative series around the open-closed background $(M,\alpha_M)$:
\be
\widetilde\Phi_{M,\alpha_M}(1/N) = N^{b_{M,\alpha_M}} \sum_{n=0}^\infty \frac{a_n^{(M,\alpha_M)}}{N^n}.
\ee
We will assume that, as is typical of the genus expansion in theories of $N \times N$ matrices, the coefficients of $\widetilde\Phi_{M,\alpha_M}$ grow as
\be
a_n^{(M,\alpha_M)} \sim n! \sim (2h)!
\ee
at genus $h$. We will refer to a large $N$ transseries of the form \eqref{eq: grav} as a \emph{gravitational transseries}. The gravitational transseries \eqref{eq: grav} is, in principle, the best we can hope to compute from the presently-known bulk path integral methods.

But to what extent does $Z_N^{\rm grav}$ furnish the actual asymptotics of $Z_N$ at large positive $N$? Or, to ask a bit more from $Z_N^{\rm grav}$, to what extent can the data contained in $Z_N^{\rm grav}$ determine the quantity $Z_N$ at finite $N$, allowing for resummation methods? These questions are motivated by two considerations:

The first is a puzzle about the structure of the large $N$ expansion. The $(2h)!$ growth of the perturbative coefficients signals, via resurgence arguments, non-perturbative sectors of order $e^{-N}$, i.e.\ D-brane instantons \cite{Shenker:1990uf}. But in holography, we are often interested in non-perturbative saddles of order $e^{-N^2}$, as these gravitational asymptotics govern the phases of the theory. How are the $e^{-N^2}$ gravity saddles encoded in the data of a large $N$ asymptotic expansion, and what do the $e^{-N^2}$ sectors have to do with the $e^{-N}$ D-brane instantons?

The second is the recent suggestion that large $N$ observables above the black hole threshold contain an ``erratic'' $N$-dependence $\delta_N$ not captured by an ordinary large $N$ expansion \cite{Schlenker:2022dyo,Kudler-Flam:2025cki,Kudler-Flam:2026nzz,Liu:2025ikq,Liu:2026fnd,Perlmutter:2026trb}. One way of stating this expectation is as the large $N$ decomposition
\be
Z_N \sim Z_N^{\rm grav} + \delta_N.
\ee
If this is true, it suggests a precision bound set by the size $|\delta_N| = \sqrt{\la |\delta_N|^2 \ra_N}$, defined via a suitable averaging over the values of $N$, beyond which $Z_N^{\rm grav}$ ceases to be a reliable estimate of $Z_N$ at finite $N$.

In this work, we will ask whether the existence of this ``precision bound'' can be seen directly from the bulk data $Z_N^{\rm grav}$. That is, if it were true that $Z_N \sim Z_N^{\rm grav} + \delta_N$, does the gravitational transseries $Z_N^{\rm grav}$ know about its own eventual failure to determine the exact microscopic answer $Z_N$?  We propose that the answer is yes.

We propose that, above the black hole threshold, an analytic obstruction prevents the resummation of $Z_N^{\rm grav}$ from determining $Z_N$ beyond a precision $e^{-N^2\hat I}$, where the precision scale $\hat I$ is set by the data of the large $N$ D-brane instanton sectors. To arrive at this proposal, we develop ``higher'' resummation methods in large $N$ resurgence that capture the $e^{-N^2}$ asymptotics of gravitational saddles, beyond the $e^{-N}$ instantons. The two scales $e^{-N}$ and $e^{-N^2}$ will correspond to singularities on Borel planes of different ``levels''. This is based on \'{E}calle's so-called \emph{accelero-summation} which, to our knowledge, has not been applied in the context of the large $N$ expansion \cite{_calle_1993,Ecalle:GuidedTour,Bellon:2018lwy}.

In Section \ref{sec: higher}, we introduce accelero-summation. In the large $N$ expansion, this allows us to treat non-perturbative effects of orders $e^{-N}$ and $e^{-N^2}$ on Borel planes of different levels.

In Section \ref{sec: bound}, under an assumption on the growth of the D-brane instanton sums that we expect to hold in the black hole phase, we find an analytic obstruction to the resummation of $Z_N^{\rm grav}$. We propose that this obstruction prevents $Z_N^{\rm grav}$ from determining $Z_N$ beyond a precision set by the data of its D-brane instanton sectors.

In Section \ref{sec: index}, we study the $\frac{1}{16}$-BPS index of $\cN=4$ super Yang-Mills and its black hole as our main example and consistency check. Using higher resummation, we derive relations between the D-brane instanton data on thermal AdS and on the black hole, which reproduce the actions of wrapped D3-branes on the $\frac{1}{16}$-BPS black hole.

In Section \ref{sec: general}, we include the higher open string corrections and generalize the precision bound beyond the annulus. We find that the D-brane instanton data on thermal AdS and on the black hole are related by a duality that exchanges the numbers of D-brane instantons on the two backgrounds.

In Section \ref{sec: generic}, we state our expectations for generic situations and discuss their relation to the double-cone wormhole. We conclude with open questions in the Discussion.

\section{Higher Borel-\'{E}calle resummation}
\label{sec: higher}

\subsection{Basic notions}
\label{subsec: basic}

Let us recall some relevant basic notions in Borel resummation and resurgence \cite{Ecalle:1981:FRTa,Ecalle:1981:FRTb,Ecalle:1985:FRTc}, with an emphasis on assumptions that are standard in familiar resurgent contexts but need not hold, as we will argue in subsequent sections, for the large $N$ expansion above the black hole threshold.

The (level-1) Borel resummation of an asymptotic series $\tilde\Phi(g)$ proceeds in two steps. The Borel transform $\cB$ maps a factorically-divergent series to one with a finite radius of convergence, defining an analytic function near the origin of the Borel plane. The Laplace integral $\cL^\theta$ then maps the analytic continuation of this function along a ray to a function $\Phi(g)$ of $g$ whose asymptotic expansion at small $g$ is the original series $\tilde\Phi(g)$. The result of this procedure depends on the choice of Laplace ray only through the singularities of the Borel transform. A singularity on the Borel plane introduces a discontinuity for the Laplace integrals taken on its either side. These Stokes discontinuities encode non-perturbative sectors that govern the small $g$ asymptotics of the physical quantity under consideration.

Let
\be
\tilde\Phi(g) = \sum_{n=0}^\infty a_n g^n, \qquad |a_n| \lesssim A^{-n} \, n!
\ee
be an asymptotic series in the parameter $g$. Its Borel transform
\be \label{eq: basic borel}
B(\zeta) \equiv (\cB \tilde\Phi) (\zeta) = \sum_{n=0}^\infty \frac{a_n}{\Gamma(n+1)} \zeta^n
\ee
converges in a disk of finite radius $A$ about the origin of the Borel $\zeta$-plane.

Let us assume that the Borel transform $B(\zeta)$ can be analytically continued along a ray at an angle $\theta$ on the Borel $\zeta$-plane and that $B(\zeta)$ does not ``grow too rapidly'' along that ray. Then the (directional) Laplace integral
\be \label{eq: basic laplace int}
(\cL^\theta B) (g) = \frac{1}{g} \int_0^{e^{i \theta} \infty} e^{-\zeta/g} \, B(\zeta) \, d\zeta
\ee
defines an actual function $\Phi(g) \equiv (\cL^\theta B) (g)$ whose asymptotic expansion around $g=0$ is the original perturbative series:
\be
\Phi(g) \sim \tilde\Phi(g).
\ee
That $\Phi(g) \sim \tilde\Phi(g)$ follows from a formal termwise application of the Laplace integral \eqref{eq: basic laplace int} to the Borel transform \eqref{eq: basic borel}. For a single monomial, we have
\be
\cL^{\theta} \left[ \frac{\zeta^{n}}{\Gamma(n+1)} \right] (g) = \frac{1}{g \, \Gamma(n+1)}\int_0^{e^{i \theta} \infty} e^{-\zeta/g} \zeta^{n} \, d\zeta = g^n.
\ee
If $g$ is positive, the integral \eqref{eq: basic laplace int} is convergent only along rays in the half-plane $-\frac{\pi}{2} < \theta < \frac{\pi}{2}$ for $\theta$ that satisfy our assumptions. We can extend this domain by compensating for the angle $\theta$ of the Laplace ray with a synchronized phase $g \to e^{i \theta} g$ for the coupling $g$. Doing so preserves the convergence of the Laplace integral, for all $\theta$ that satisfy our assumptions, and will make manifest the connection between the Laplace integral as a function of $\theta$ and the Stokes transition that occurs as a function of the complex phase of $g$.

When we say that $B(\zeta)$ does not ``grow too rapidly'' along the ray, we mean that it grows slowly enough for the Laplace integral \eqref{eq: basic laplace int} to converge:
\be \label{eq: basic laplace bound}
\int_0^{e^{i\theta}\infty} \big| e^{-\zeta/g}\, B(\zeta) \big| \, |d\zeta| < \infty.
\ee
Thus, an asymptotic series $\tilde\Phi(g)$ is Borel summable along $\theta$ \emph{using the methods of this subsection} if its Borel transform $B(\zeta)$ satisfies \eqref{eq: basic laplace bound} for sufficiently small $g$. This is the case if $B(\zeta)$ grows at most exponentially along the ray. While this growth condition is satisfied for the Borel transforms that appear in familiar resurgent contexts, we will argue that precisely this assumption fails for the large $N$ expansion of observables above the black hole threshold.

A singularity on the Borel $\zeta$-plane indicates a non-perturbative sector $e^{-A/g}$ of a physical observable (that may contribute, depending on additional physical input, to its small $g$ asymptotics) in the following sense. Let the Borel transform $B(\zeta)$ have an isolated logarithmic singularity at $\zeta = A$. It may also have isolated singularities elsewhere. Then we can expand $B(\zeta)$ locally near the singularity as 
\be \label{eq: basic log sing}
B(\zeta) = \frac{i}{2 \pi} \log(\zeta - A) \sum_{n=0}^\infty \frac{c_n}{\Gamma(n+1)} (\zeta - A)^n \, + \, \text{regular}.
\ee
The presence of this singularity affects the output of the Laplace integral as we rotate its ray counterclockwise on the Borel plane across the angle $\theta = \arg A$. The singularity \eqref{eq: basic log sing} contributes, to the difference
\be
(\cL^{\theta^+} B-\cL^{\theta^-} B ) (g)
\ee
between the Laplace integrals across direction $\theta$, as
\be \label{eq: basic log discontinuity}
\frac{1}{g} \int_{\cK_A} e^{-\zeta/g} \, B(\zeta) \, d\zeta \, = \, e^{-A/g} \sum_{n=0}^\infty c_n  g^n.
\ee
Here, the contour for the difference between the two rays was deformed into the clockwise Hankel contour $\cK_{A}$ surrounding a logarithmic branch cut that ends at the branch point $\zeta=A$. This is the contribution of \eqref{eq: basic log sing} to the Stokes discontinuity across $\theta = \arg A$.\footnote{Typically, a Stokes discontinuity refers to an abrupt change, by a subleading exponential, in the asymptotic behavior of a function as we rotate the phase of $g$, as opposed to the rotation of the angle $\theta$ associated with Borel resummation. The two become equivalent if we, as explained above, synchronize the phase of the coupling as $g \to e^{i \theta} g$. Therefore, a discontinuity in the Laplace integral at angle $\theta$ corresponds to a Stokes discontinuity at angle $\theta$ on the complex $g$-plane.}

If the Borel transform $B(\zeta)$ had multiple singularities along $\theta = \arg A$, e.g, for multi-instantons
\be \label{eq: basic multi log sing}
B(\zeta) = \frac{i}{2 \pi} \sum_{\ell=1}^\infty \left( \log(\zeta - \ell A) \sum_{n=0}^\infty \frac{c_n^{(\ell)}}{\Gamma(n+1)} (\zeta - \ell A)^n \, + \, \text{regular} \right),
\ee
the Stokes discontinuity is given by a sum over instanton sectors arising from each singularity:
\ie \label{eq: basic multi log discontinuity}
(\cL^{\theta^+} B - \cL^{\theta^-} B) (g) &= \frac{1}{g} \sum_{\ell=1}^\infty \int_{\cK_{\ell A}} e^{-\zeta/g} \, B(\zeta) \, d\zeta \\
&= \sum_{\ell=1}^\infty e^{-\ell A/g} \sum_{n=0}^\infty c_n^{(\ell)}  g^n. 
\fe
In the first line, the contour for the difference between the two rays was deformed into a sum over the clockwise Hankel contours $\cK_{\ell A}$.

To conclude, let us comment on how the Borel resummation described here is related to the small $g$ asymptotics of a physical observable $\cO(g)$. Once the paths of analytic continuation of $B(\zeta)$ on its Riemann surface are specified, the difference $(\cL^{\theta^+} B - \cL^{\theta^-} B)(g)$ across a Stokes direction $\theta$ is unambiguous provided the Laplace integrals exist. To determine the coefficients with which the non-perturbative sectors contribute to the small $g$ asymptotics of $\cO(g)$, however, we require additional physical input such as an integral definition for $\cO(g)$ together with its contour. The Borel singularities nevertheless determine the form of the non-perturbative contributions that can enter via Stokes transitions.

\subsection{Higher resummation and acceleration}
\label{subsec: acceleration}

The preceding discussion assumed that the Borel transform admits the analytic continuation and growth property \eqref{eq: basic laplace bound} required for the (level-1) Laplace integrals to exist. We will now discuss how Borel resummation may be extended when the growth condition fails, using an operation known as acceleration \cite{_calle_1993,Ecalle:GuidedTour,AIHPA_1991__54_4_331_0}.

Our motivation is the following. The large $N$ expansion of a theory of $N \times N$ matrices has coefficients that grow as $n! = (2h)!$ at genus $h$, which indicates the presence of $e^{-N}$ D-brane instanton sectors. In holographic contexts, however, we are often interested in ``gravitational'' contributions of order $e^{-N^2}$ since these asymptotics control the thermodynamic phases of the theory. We will show that the $e^{-N^2}$ sectors are associated with the behavior of $B(\zeta)$ at large $\zeta$, which was constrained by the growth condition \eqref{eq: basic laplace bound} in the preceding discussion, and that they appear as singularities on a ``level-2 Borel plane'' obtained from the level-1 Borel $\zeta$-plane via an integral transform known as acceleration.

\subsubsection*{Level-$k$ Borel-\'{E}calle resummation}

Let
\be
\tilde\Phi(g) = \sum_{n=0}^\infty a_n g^n, \qquad |a_n| \lesssim A^{-n} \, n!
\ee
be an asymptotic series as in Section \ref{subsec: basic}. We will, however, relax the growth condition \eqref{eq: basic laplace bound} of the level-1 Borel transform $B(\zeta)$ at large $\zeta$.

For $k>0$, the level-$k$ Borel transform of $\tilde\Phi(g)$ is
\be \label{eq: borel defn}
B_k(\zeta_k) \equiv (\cB_k \tilde\Phi) (\zeta_k) = \sum_{n=0}^\infty \frac{a_n}{\Gamma(n/k + 1)} \zeta_k^{n/k},
\ee
and the level-$k$ (directional) Laplace integral along a ray at angle $\theta_k$ on the Borel $\zeta_k$-plane is
\be \label{eq: laplace defn}
(\cL_k^{\theta_k} B_k) (g) = \frac{1}{g^k} \int_0^{e^{i \theta_k} \infty} e^{-\zeta_k/g^k} \, B_k(\zeta_k)\, d\zeta_k.
\ee
For $k=1$, these reduce to \eqref{eq: basic borel} and \eqref{eq: basic laplace int}, and we will use $\zeta_1 \equiv \zeta$, $B_1 \equiv B$, and $\cL_1^{\theta_1} \equiv \cL^\theta$ interchangeably. For a single monomial, we have
\be
\cL_k^{\theta_k} \left[ \frac{\zeta_k^{n/k}}{\Gamma(n/k+1)} \right] (g) = \frac{1}{g^k \, \Gamma(n/k+1)}\int_0^{e^{i \theta_k} \infty} e^{-\zeta_k/g^k} \zeta_k^{n/k} \, d\zeta_k = g^n,
\ee
so a termwise application of the Laplace integral to $B_k(\zeta_k)$ returns $\tilde\Phi(g)$. When it exists, we call $(\cL_k^{\theta_k} B_k) (g)$ the level-$k$ Borel resummation of $\tilde\Phi(g)$.\footnote{Our normalization of the Borel transform \eqref{eq: borel defn} differs from that of \'{E}calle \cite{Ecalle:GuidedTour}, in which, for $n>0$, $g^n \mapsto \zeta_k^{n/k-1}/\Gamma(n/k)$. Our Borel transform is the $\zeta_k$-primitive of \'{E}calle's, up to a constant term. Correspondingly, the Laplace integral in \cite{Ecalle:GuidedTour} lacks the prefactor $g^{-k}$ of \eqref{eq: laplace defn}. \label{foot: conv}}

Level-$k$ Borel resummation is simply level-1 Borel resummation in the variable $g^k$, so the discussion of Section \ref{subsec: basic} carries through with $\zeta/g \to \zeta_k/g^k$. In particular, we compensate for the angle $\theta_k$ of the Laplace ray on the $\zeta_k$-plane with a synchronized phase $g \to e^{i\theta} g$ for the coupling, where $\theta_k = k\theta$. This is so that the Laplace kernel $e^{-\zeta_k/g^k}$ decays along the ray for every $\theta_k$. The condition \eqref{eq: basic laplace bound} then becomes the requirement that the level-$k$ Laplace integral converges:
\be \label{eq: level k laplace bound}
\int_0^{e^{i\theta_k}\infty} \big| e^{-\zeta_k/g^k}\, B_k(\zeta_k) \big| \, |d\zeta_k| < \infty.
\ee
As before, this holds if $B_k(\zeta_k)$ grows at most exponentially along the ray (for sufficiently small $g$).

To characterize singularities on the $\zeta_k$-plane, it is convenient to use the discontinuity
\be
\Disc_{\theta_k} B_k(\zeta_k) = B_k^{\theta_k^+}(\zeta_k) - B_k^{\theta_k^-}(\zeta_k), 
\ee
between the analytic continuations of $B_k(\zeta_k)$ along the rays at angles $\theta_k^\pm$ just counterclockwise and clockwise of $\theta_k$, evaluated on the ray at angle $\theta_k$. For example, the discontinuity of the logarithmic singularity \eqref{eq: basic log sing} across $\theta_1 = \arg A$ is
\be
\Disc_{\theta_1} B(\zeta) = \Theta(|\zeta| - |A|) \sum_{n=0}^\infty \frac{c_n}{\Gamma(n+1)} (\zeta - A)^n,
\ee
where $\Theta$ denotes the Heaviside function. The integral along the Hankel contour $\cK_A$ in \eqref{eq: basic log discontinuity} becomes the Laplace integral of this discontinuity along the cut starting at $\zeta = A$.

More generally, consider a level-$k$ Borel transform with a singularity at $\zeta_k = S$ whose discontinuity across $\theta_k = \arg S$ is
\be \label{eq: general sing}
\Disc_{\theta_k} B_k(\zeta_k) = \Theta(|\zeta_k| - |S|) \sum_{n=0}^\infty \frac{c_n}{\Gamma(n/k + 1 - \nu)} \, (\zeta_k - S)^{n/k - \nu},
\ee
where we fix $\nu<1$ by requiring that $c_0$ is non-vanishing. The case $k = 1$ and $\nu = 0$ corresponds to the logarithmic singularity \eqref{eq: basic log sing}.\footnote{When $\nu = 0$ and only the coefficients $c_n$ with $n \in k\bZ_{\geq 0}$ are non-vanishing, \eqref{eq: general sing} is a logarithmic discontinuity multiplying integer powers of $\zeta_k - S$ as in \eqref{eq: basic log sing}.} In analogy to the computation in Section \ref{subsec: basic}, the singularity \eqref{eq: general sing} contributes to the difference between the Laplace integrals \eqref{eq: laplace defn} across $\theta_k = \arg S$ as
\be \label{eq: general discontinuity}
\frac{1}{g^k} \int_S^{e^{i \theta_k} \infty} e^{-\zeta_k/g^k} \, \Disc_{\theta_k} B_k(\zeta_k) \, d\zeta_k = e^{-S/g^k} \, g^{-k\nu} \sum_{n=0}^\infty c_n \, g^n.
\ee
Therefore, a singularity on the Borel $\zeta_k$-plane and its local data translates into a non-perturbative sector of the form $e^{-S/g^k}$ and the perturbative fluctuations in $g$ around that sector. For the large $N$ expansion with $g = 1/N$, a singularity \eqref{eq: general sing} at $\zeta_2 = S$ on the level-2 Borel plane would translate into the term
\be
e^{-N^2 S} N^{2\nu} \sum_{n=0}^\infty \frac{c_n}{N^n}
\ee
that may contribute, depending on additional physical input, to the asymptotics of a physical observable $\cO_N$ at large positive $N$.

There is, however, an important difference between $k=1$ and $k>1$. If the coefficients $a_n$ of $\tilde\Phi(g)$ grow as $a_n \sim A^{-n} \, n!$, the ordinary Borel transform $B_1(\zeta_1)$ has a finite radius of convergence. On the other hand, for $k>1$ the coefficients $a_n/\Gamma(n/k+1)$ of \eqref{eq: borel defn} still grow factorially. So $B_k(\zeta_k)$ actually has a vanishing radius of convergence and \eqref{eq: borel defn} is only a formal power series in $\zeta_k$. It turns out, nevertheless, that we can obtain $B_k(\zeta_k)$ as an actual function in $\zeta_k$ via an integral transform of the analytic continuation of $B_1(\zeta_1)$. This operation is known as acceleration.

\subsubsection*{Accelero-summation}

Let $0 < k < k'$. The acceleration operator $\cA^{\theta_k}_{k \to k'}$ maps a Borel transform on the $\zeta_k$-plane to one on the $\zeta_{k'}$-plane. It is defined as
\be \label{eq: acc defn}
(\cA_{k \to k'}^{\theta_k} B_k)(\zeta_{k'}) = \int_{0}^{e^{i \theta_k}\infty} K_{\alpha}(\zeta_{k'},\zeta_k) \, B_k(\zeta_k) \, d\zeta_k,
\ee
where $\alpha = k/k'$. The acceleration kernel is defined as\footnote{For the same reason as in footnote \ref{foot: conv}, \'{E}calle's minor acceleration kernel \cite{Ecalle:GuidedTour} is the inverse Laplace transform of $e^{-\zeta_k x^\alpha}$ rather than of $x^{\alpha-1} e^{-\zeta_k x^\alpha}$, i.e.\ it lacks the factor $x^{\alpha-1}$ in \eqref{eq: acc kernel}.}
\be \label{eq: acc kernel}
K_{\alpha}(\zeta_{k'},\zeta_k) = \frac{1}{2 \pi i} \int_{c-i \infty}^{c+i \infty} x^{\alpha-1} \exp \! \left( x \zeta_{k'} - \zeta_k x^\alpha \right) dx, \qquad c>0.
\ee
Let us note some properties of the acceleration kernel that will be relevant for us:
\begin{enumerate}
\item $K_\alpha(\zeta_{k'},\zeta_k)$ is the inverse Laplace transform of $x^{\alpha-1} e^{-\zeta_k x^\alpha}$ with respect to the variable $x$. Inverting that relation, the acceleration kernel $K_\alpha$ satisfies the forward Laplace transform
\be \label{eq: kernel laplace}
\int_0^\infty e^{-x \zeta_{k'}} \, K_\alpha(\zeta_{k'},\zeta_k) \, d\zeta_{k'} = x^{\alpha-1} e^{-\zeta_k x^\alpha}
\ee
for $\re x > 0$ and $\zeta_k > 0$.
\item The kernel has the homogeneity property
\be \label{eq: kernel homogeneity}
K_\alpha(\lambda^{k'} \zeta_{k'}, \lambda^k \zeta_k) = \lambda^{-k} \, K_\alpha(\zeta_{k'},\zeta_k),
\ee
which can be seen by also rescaling the integration variable as $x \to \lambda^{-k'} x$ in \eqref{eq: acc kernel}.
\end{enumerate}

From the Laplace relation \eqref{eq: kernel laplace}, we see that the acceleration kernel is designed to relate the Laplace integrals at levels $k$ and $k'$. More explicitly, set $x = g^{-k'}$ with $g>0$ and then divide by $g^{k'}$ in \eqref{eq: kernel laplace}. We find
\be \label{eq: kernel identity}
\frac{1}{g^k} \, e^{-\zeta_k/g^k} = \frac{1}{g^{k'}} \int_0^{e^{i\theta_{k'}}\infty} e^{-\zeta_{k'}/g^{k'}} \, K_{\alpha}(\zeta_{k'},\zeta_k) \, d\zeta_{k'},
\ee
initially for $\zeta_k > 0$ and $\theta_{k'} = 0$. But, using the homogeneity property \eqref{eq: kernel homogeneity} with $\lambda = e^{i\theta}$, the identity \eqref{eq: kernel identity} continues to hold for complex $g$ with $\theta = \arg g$, provided that we take $\zeta_k$ to lie on the ray at angle $\theta_k = k\theta$ and that we integrate along the ray at angle $\theta_{k'} = k'\theta$. (This shows why it is natural to synchronize the phase $e^{i\theta}$ of $g$ with that $e^{i \theta_k}$ in the Laplace integral as $\theta_k = k \theta$, as discussed above \eqref{eq: level k laplace bound}.)

Substituting \eqref{eq: kernel identity} for the Laplace kernel in the level-$k$ Borel resummation \eqref{eq: laplace defn}, we have
\ie \label{eq: acc exchange}
(\cL_k^{\theta_k} B_k) (g) &= \int_0^{e^{i\theta_k}\infty} \left( \frac{1}{g^{k'}} \int_0^{e^{i\theta_{k'}}\infty} e^{-\zeta_{k'}/g^{k'}} \, K_\alpha(\zeta_{k'},\zeta_k) \, d\zeta_{k'} \right) B_k(\zeta_k) \, d\zeta_k \\
&= \frac{1}{g^{k'}} \int_0^{e^{i\theta_{k'}}\infty} e^{-\zeta_{k'}/g^{k'}} \, (\cA^{\theta_k}_{k \to k'} B_k)(\zeta_{k'}) \, d\zeta_{k'}.
\fe
In the second line, we exchanged the order of the $\zeta_k$ and $\zeta_{k'}$ integrals, and we will comment on this shortly. The result is the level-$k'$ Laplace integral \eqref{eq: laplace defn} of $\cA^{\theta_k}_{k \to k'} B_k$. Therefore,
\be \label{eq: acc equality 2}
\cL_{k'}^{\theta_{k'}} \, \cA^{\theta_k}_{k \to k'} B_k = \cL_k^{\theta_k} B_k.
\ee
If $B_k(\zeta_k)$ satisfies the growth condition \eqref{eq: level k laplace bound}, the exchange of the $\zeta_k$ and $\zeta_{k'}$ integrals in \eqref{eq: acc exchange} is justified and both sides of \eqref{eq: acc equality 2} exist and agree.\footnote{By Fubini, the exchange is justified if the double integral of the absolute value of the integrand converges.}

The situation of interest to us, however, will be one in which the RHS does not exist because $B_k(\zeta_k)$ grows too rapidly for the level-$k$ Laplace integral to converge. We will therefore use the LHS of \eqref{eq: acc equality 2} to define the Borel resummation of $\tilde\Phi(g)$ in such situations. The level $k'$ must be chosen so that it is small enough for the acceleration integral \eqref{eq: acc defn} to converge but large enough for the subsequent level-$k'$ Laplace integral to exist. As we will see below, this fixes a critical value of $k'$ set by the growth of $B_k(\zeta_k)$.

We can check that the acceleration $\cA^{\theta_k}_{k \to k'}$ of the level-$k$ Borel transform $B_k(\zeta_k)$ is compatible with the definition \eqref{eq: borel defn} of the level-$k'$ Borel transform $B_{k'}(\zeta_{k'})$. On monomials, it acts by \eqref{eq: kernel laplace} and \eqref{eq: kernel homogeneity} as
\be \label{eq: acc monomial}
\cA^{\theta_k}_{k \to k'} \left[ \frac{\zeta_k^{n/k}}{\Gamma(n/k +1)} \right] (\zeta_{k'}) = \frac{\zeta_{k'}^{n/k'}}{\Gamma(n/k' +1)}.
\ee
for $\zeta_{k'}$ on the ray at angle $\theta_{k'}$. Applying this termwise to \eqref{eq: borel defn}, we see that acceleration maps the level-$k$ Borel transform of $\tilde\Phi(g)$ to its level-$k'$ Borel transform:
\be \label{eq: acc equality 1}
\cA_{k \to k'}^{\theta_k} B_k = B_{k'}.
\ee
The RHS of \eqref{eq: acc equality 1} is only a formal series for $k'>1$, so this termwise equality is formal. The LHS, however, defines an actual function of $\zeta_{k'}$ (for a suitable choice of $k'$) as it is an integral transform of the analytic continuation of $B_k(\zeta_k)$ along $\theta_k$. We therefore take the LHS of \eqref{eq: acc equality 1} as the definition of the level-$k'$ Borel transform $B_{k'}(\zeta_{k'})$ as an actual function of $\zeta_{k'}$. Unlike the formal series on the RHS, the function on the LHS depends on the ray $\theta_k$ and thus changes discontinuously when the ray crosses a singularity of $B_k(\zeta_k)$.

\subsubsection*{Critical level $k'$}

We will now determine the critical level $k'$ to which to accelerate. Suppose that the level-$k$ Borel transform grows as
\be \label{eq: B1 growth}
\log | B_k(\zeta_k) | \sim C \, |\zeta_k|^p, \qquad p>1, \quad C>0
\ee
as $\zeta_k \to e^{i\theta_k} \infty$, so that the level-$k$ Borel resummation does not exist for any $g$. As we stated above, the level $k'$ is required to be (1) small enough for the acceleration integral to converge but (2) large enough for the subsequent level-$k'$ Laplace integral to exist.

Whether the acceleration integral \eqref{eq: acc defn} converges depends on the behavior of its integrand at large $\zeta_k$. A saddle point evaluation of the kernel \eqref{eq: acc kernel} as $\zeta_k \to e^{i \theta_k} \infty$ gives
\be \label{eq: kernel saddle}
K_\alpha(\zeta_{k'},\zeta_k) \sim \exp \! \left[ - (1-\alpha) \, \alpha^{\frac{\alpha}{1-\alpha}} \, \zeta_k^{\frac{k'}{k'-k}} \, \zeta_{k'}^{-\frac{k}{k'-k}} \right]
\ee
at fixed $\zeta_{k'}$. The exponent is real and negative for $\zeta_{k'}$ on the synchronized ray at angle $\theta_{k'}$. Let us write $q \equiv \frac{k'}{k'-k}$. Comparing \eqref{eq: B1 growth} and \eqref{eq: kernel saddle}, we see that the acceleration integral diverges for all $\zeta_{k'}$ if $q < p$, while it converges for all $\zeta_{k'}$ along (and in a neighborhood of) the ray if $q > p$. In the marginal case $q = p$, the growth of $B_k(\zeta_k)$ and the decay of the kernel are of the same order, so whether the acceleration integral converges depends on the value of $\zeta_{k'}$. To satisfy requirement (1), $k'$ must thus be chosen so that $q \geq p$.

Suppose now that $q > p$ so that the acceleration integral converges for $\zeta_{k'}$ near its synchronized ray. By \eqref{eq: B1 growth} and \eqref{eq: kernel saddle}, the magnitude of the integrand behaves at large $|\zeta_k|$ as
\be
\exp \! \left[ C |\zeta_k|^p - (1-\alpha)\,\alpha^{\frac{\alpha}{1-\alpha}} |\zeta_k|^q |\zeta_{k'}|^{1-q} \right]
\ee
for $\zeta_{k'}$ on the synchronized ray. This is maximized at $|\zeta_k| \propto |\zeta_{k'}|^{\frac{q-1}{q-p}}$, which grows with $|\zeta_{k'}|$. A saddle point evaluation of the acceleration integral then gives
\be
\log |B_{k'}(\zeta_{k'})| \sim C' |\zeta_{k'}|^{\frac{p(q-1)}{q-p}}
\ee
for some $C'>0$, where $\frac{p(q-1)}{q-p} > 1$. The level-$k'$ Laplace integral therefore diverges, and the case $q > p$ does not satisfy requirement (2).

What about the marginal case $q = p$? Now the magnitude of the acceleration integrand behaves at large $|\zeta_k|$ as
\be
\exp \! \left[ - \left( (1-\alpha)\,\alpha^{\frac{\alpha}{1-\alpha}} |\zeta_{k'}|^{1-p} - C \right) |\zeta_k|^p \right],
\ee
so the acceleration integral converges for $\zeta_{k'}$ on the synchronized ray with
\be
|\zeta_{k'}| < \left( (1-\alpha)\,\alpha^{\frac{\alpha}{1-\alpha}} C^{-1} \right)^{\frac{1}{p-1}}.
\ee
(Unlike for $q > p$, the integrand here has no saddle that moves out to infinity as $|\zeta_{k'}|$ grows, so $B_{k'}(\zeta_{k'})$ is not forced to grow faster than exponentially.) Beyond this bound, $B_{k'}(\zeta_{k'})$ is no longer given by the acceleration integral, so the level-$k'$ Laplace integral requires the analytic continuation of $B_{k'}(\zeta_{k'})$ along the rest of the ray. If this continuation of $B_{k'}(\zeta_{k'})$ exists and grows at most exponentially in $|\zeta_{k'}|$, then the level-$k'$ Laplace integral would converge for sufficiently small $g$ in the sense of \eqref{eq: level k laplace bound}.\footnote{If the continuation has singularities on the ray, we instead take the Laplace integrals along the rays at angles $\theta_{k'}^\pm$ deformed to either side.} We are not aware of general conditions on $B_k(\zeta_k)$ that guarantee these properties for the continuation of $B_{k'}(\zeta_{k'})$. We will instead examine whether this holds for the Borel transforms at hand.

To summarize, requirement (1) excludes $q < p$ and (2) excludes $q > p$, so that only the marginal case $q = p$ can satisfy both. The critical level $k'$ is therefore given by
\be \label{eq: critical level}
\frac{1}{k'} = \frac{1}{k} \left( 1 - \frac{1}{p} \right).
\ee
We will illustrate in Section \ref{subsec: level 12} how this analytic continuation works in a simple example, and show in Section \ref{sec: bound} that it can be obstructed.

\subsection{Acceleration from level 1 to level 2}
\label{subsec: level 12}

Let us now specialize to $k = 1$ and $k' = 2$, as they will be the levels relevant for the large $N$ transseries. The level-1 singularities correspond to the $e^{-N}$ D-brane instanton sectors and the level-2 singularities to the $e^{-N^2}$ gravitational sectors. We set $g = 1/N$, so that the phase of the coupling is $\theta = \arg 1/N$. The Laplace integrals at levels 1 and 2 read
\ie \label{eq: laplace N}
(\cL_1^{\theta_1} B_1)(1/N) &= N \int_0^{e^{i\theta_1}\infty} e^{-N\zeta_1} \, B_1(\zeta_1) \, d\zeta_1 \\
(\cL_2^{\theta_2} B_2)(1/N) &= N^2 \int_0^{e^{i\theta_2}\infty} e^{-N^2\zeta_2} \, B_2(\zeta_2) \, d\zeta_2,
\fe
with $\theta_1 = \theta$ and $\theta_2 = 2\theta$.

For these levels, the acceleration kernel is a gaussian in $\zeta_1$:
\be \label{eq: kernel N}
K_{1/2}(\zeta_2,\zeta_1) = \frac{1}{\sqrt{\pi \zeta_2}} \exp \! \left( -\frac{\zeta_1^2}{4 \zeta_2} \right).
\ee
The level-2 Borel transform is
\be \label{eq: acc 12}
B_2(\zeta_2) \equiv (\cA^{\theta_1}_{1 \to 2} B_1)(\zeta_2) = \frac{1}{\sqrt{\pi \zeta_2}} \int_0^{e^{i\theta_1}\infty}  e^{-\frac{\zeta_1^2}{4 \zeta_2}} B_1(\zeta_1) \, d\zeta_1,
\ee
and the identity \eqref{eq: kernel identity} reads
\be \label{eq: kernel identity N}
N e^{-N\zeta_1} = N^2 \int_0^{e^{i\theta_2}\infty}  \frac{1}{\sqrt{\pi\zeta_2}} \, e^{-\frac{\zeta_1^2}{4\zeta_2}} \, e^{-N^2\zeta_2} \, d\zeta_2
\ee
for $\zeta_1$ on the ray at angle $\theta_1$, where $\theta_2 = 2\theta_1$ is the direction of the level-2 Laplace ray. On the other hand, note that $\zeta_2$ is a free variable in the acceleration integral \eqref{eq: acc 12} subject only to the condition that the integral should converge.

In Section \ref{sec: bound}, we will find that if the one-loop factors $e^{-\ell^2 \gamma}$ of the $\ell$-instanton sectors $e^{-\ell N A}$ grow rather than decay with $\ell$, i.e.\ if $\re \gamma < 0$, the discontinuity of $B_1(\zeta_1)$ across $\theta_1 = \arg A$ grows as a gaussian. Let us therefore consider a level-1 Borel transform (or, as will be the case below, its discontinuity) that grows as
\be \label{eq: B1 growth N}
\log | B_1(\zeta_1) | \sim \frac{1}{4\rho} |\zeta_1|^2 \ \ \text{as} \ \ \zeta_1 \to e^{i \theta_1} \infty, \qquad \rho>0.
\ee
This is the growth \eqref{eq: B1 growth} with $p=2$ and $C = \frac{1}{4 \rho}$, so $k'=2$ is indeed the critical level \eqref{eq: critical level}. Let us write $\zeta_1 = r_1 e^{i\theta_1}$ along the integration ray of \eqref{eq: acc 12}. At large $r_1$, the integrand of \eqref{eq: acc 12} then behaves as
\be
\exp \! \left[ - \frac{r_1^2}{4} \left( \re \! \left( \frac{e^{2i\theta_1}}{\zeta_2} \right) - \frac{1}{\rho} \right) \right],
\ee
so the acceleration integral converges for $\re(e^{2i\theta_1}/\zeta_2) > 1/\rho$. That is, \eqref{eq: acc 12} defines $B_2(\zeta_2)$ in the disk
\be \label{eq: acc disk}
\left| \zeta_2 - \frac{\rho}{2} \, e^{2i\theta_1} \right| < \frac{\rho}{2}.
\ee
The disk depends only on $\rho$ and on the acceleration angle $\theta_1$ (but not on $\zeta_2$). Its boundary circle passes through the origin and crosses the level-2 Laplace ray at $\zeta_2 = \rho\, e^{i\theta_2}$. The level-2 Borel resummation requires the analytic continuation of $B_2(\zeta_2)$ beyond that point.

\subsubsection*{Gaussian toy growth}

Let us illustrate how this analytic continuation works in the toy Borel transform
\be
B_1(\zeta_1) = e^{\zeta_1^2/4\rho}
\ee
with $\theta = 0$. The level-1 Laplace integral of $B_1(\zeta_1)$ diverges for every $N$. But the acceleration integral \eqref{eq: acc 12} converges in the disk \eqref{eq: acc disk} and gives
\be \label{eq: gaussian example}
B_2(\zeta_2) = \frac{1}{\sqrt{1 - \zeta_2/\rho}}.
\ee
(The formal series \eqref{eq: acc equality 1} in this case converges to \eqref{eq: gaussian example} for $|\zeta_2| < \rho$.) This function continues analytically to the $\zeta_2$-plane with a cut along $[\rho, \infty)$, with a square-root branch point at $\zeta_2 = \rho$ on the boundary of the convergence disk. Its discontinuity across $\theta_2 = 0$ is
\be
\Disc_0 B_2(\zeta_2) = \Theta(\zeta_2 - \rho) \, \frac{2 i \sqrt{\rho}}{\sqrt{\zeta_2 - \rho}},
\ee
which is of the form \eqref{eq: general sing} with $k = 2$, $S = \rho$, $\nu = \frac{1}{2}$, $c_0 = 2 i \sqrt{\pi \rho}$, and $c_{n>0} = 0$. The level-2 Borel resummations along $\theta_2 = 0^\pm$ therefore exist. By \eqref{eq: general discontinuity} they differ by
\be
(\cL_2^{0^+} B_2 - \cL_2^{0^-} B_2)(1/N) = 2 i \sqrt{\pi \rho} \, N e^{- N^2 \rho}.
\ee
We can interpret this as the contribution of the saddle point $\zeta_1 = 2 N \rho$ (as opposed to a singularity) of the divergent level-1 Laplace integral. On the $\zeta_2$-plane, this is represented by an ordinary singularity at finite distance $\zeta_2 = \rho$. Thus the gaussian growth of $B_1(\zeta_1)$ at large $\zeta_1$ gives rise to a singularity of $B_2(\zeta_2)$, which in turn gives rise to an $e^{-N^2}$ Stokes discontinuity.

\subsubsection*{Stokes discontinuities on the $\zeta_2$-plane}

In subsequent sections, the level-1 data will be given in the form of a Stokes discontinuity across a ray on which $B_1(\zeta_1)$ has infinitely many singularities. For example, for the multi-instanton singularities along $\theta_1 = \arg A$,
\be \label{eq: multi log sing 2}
B_1(\zeta_1) = \frac{i}{2 \pi} \sum_{\ell=1}^\infty \left( \log(\zeta_1 - \ell A) \sum_{n=0}^\infty \frac{c_n^{(\ell)}}{\Gamma(n+1)} (\zeta_1 - \ell A)^n \, + \, \text{regular} \right),
\ee
we have
\be \label{eq: multi disc}
\Disc_{\theta_1} B_1(\zeta_1) = \sum_{\ell=1}^\infty \Theta(|\zeta_1| - \ell|A|) \sum_{n=0}^\infty \frac{c_n^{(\ell)}}{\Gamma(n+1)} (\zeta_1 - \ell A)^n.
\ee
Its Laplace integral $\cL_{1}^{\theta_1} \Disc_{\theta_1} B_1$, if it exists, would give the sum over D-brane instanton sectors \eqref{eq: basic multi log discontinuity} on a fixed gravity background.

When $B_1(\zeta_1)$ is well-defined and its level-1 Laplace integrals exist, the identity $\cL_{2}^{\theta_{2}} \cA^{\theta_1}_{1 \to 2} B_1 = \cL_1^{\theta_1} B_1$ in \eqref{eq: acc equality 2} and the linearity of acceleration allow us to represent the Stokes discontinuity on the $\zeta_2$-plane as
\ie \label{eq: acc stokes}
\cL_1^{\theta_1^+} B_1 - \cL_1^{\theta_1^-} B_1 &= \cL_1^{\theta_1} \Disc_{\theta_1} B_1 \\
&= \cL_2^{\theta_2} \cA^{\theta_1}_{1\to 2} \Disc_{\theta_1} B_1
\fe
where the discontinuity of $B_2(\zeta_2)$ across $\theta_2 = 2\theta_1$ is
\be \label{eq: disc 2}
\Disc_{\theta_2} B_2 = \cA^{\theta_1^+}_{1\to 2} B_1 - \cA^{\theta_1^-}_{1\to 2} B_1 = \cA^{\theta_1}_{1\to 2} \Disc_{\theta_1} B_1.
\ee

The situation of interest to us, however, is one in which the Stokes data of the level-1 Borel singularities grow as $c_0^{(\ell)} \sim e^{-\ell^2 \gamma}$ with $\re \gamma < 0$. This will later have the consequence that the corresponding discontinuity of the level-1 Borel transform \eqref{eq: multi disc} grows as a gaussian as in \eqref{eq: B1 growth N}. (If instead $\re \gamma > 0$, acceleration is not required. But the identity \eqref{eq: acc stokes} continues to hold and we can still apply it. Later, we will find it instructive to apply acceleration even in such situations.) In this case, only some of the expressions above remain meaningful. Since the sum over $\ell$ in \eqref{eq: multi log sing 2} diverges, $B_1(\zeta_1)$ is not defined as a naive sum over its singularities. By contrast, its discontinuity $\Disc_{\theta_1} B_1(\zeta_1)$ in \eqref{eq: multi disc} is a finite sum at each point on the ray and its acceleration $\Disc_{\theta_2} B_2(\zeta_2)$ converges in the disk \eqref{eq: acc disk}.

We will therefore define the Stokes discontinuity across the ray at angle $\theta_1 = \arg A$ on the $\zeta_1$-plane, or equivalently across the ray at angle $\theta_2 = 2\theta_1$ on the $\zeta_2$-plane, as
\be \label{eq: acc stokes defn}
(\cL_2^{\theta_2} \Disc_{\theta_2} B_2) (1/N).
\ee
This is well-defined if $\Disc_{\theta_2} B_2$ can be analytically continued in $\zeta_2$ along the ray at angle $\theta_2$ with at most exponential growth. The definition \eqref{eq: acc stokes defn} agrees with $(\cL_1^{\theta_1} \Disc_{\theta_1} B_1) (1/N)$ whenever the latter exists.\footnote{We should note that $\Disc_{\theta_2} B_2 = \cA^{\theta_1}_{1\to 2} \Disc_{\theta_1} B_1$ does not strictly speaking determine $\cA^{\theta_1^\pm}_{1\to 2} B_1$. We can only say that an obstruction to the analytic continuation of $\cA^{\theta_1}_{1\to 2}\Disc_{\theta_1} B_1$ implies an obstruction for at least one of $\cA^{\theta_1^\pm}_{1\to 2} B_1$. Still, anticipating Section \ref{sec: bound}, if $\Disc_{\theta_2} B_2$ has a natural boundary so will $\cA^{\theta_1^+}_{1\to 2} B_1 + \sigma \, \Disc_{\theta_2} B_2$ for all but at most one value of $\sigma$.} For a single instanton sector, acceleration along $\theta_1 = \arg A$ gives
\be \label{eq: acc erfc}
\cA^{\theta_1}_{1\to 2} \big[ \Theta(|\zeta_1| - |A|) \big] (\zeta_2) = \erfc \! \left( \frac{A}{2\sqrt{\zeta_2}} \right).
\ee
Its level-2 Laplace integral indeed returns $e^{-NA}$. The $\erfc$ function behaves near the origin as $e^{-A^2/4\zeta_2}$, so an isolated D-brane instanton sector is represented by an essential singularity at the origin of the $\zeta_2$-plane. As in the gaussian toy example, singularities away from the origin arise from the growth of the discontinuity at large $\zeta_1$, i.e.\ from the collective contribution of the instanton sectors with large $\ell$. We will find in Section \ref{sec: bound} that these can result in a natural boundary on the $\zeta_2$-plane beyond which the level-2 Laplace integral cannot be extended.

\section{A precision bound for gravitational transseries}
\label{sec: bound}

\subsection{Large $N$ instantons above the black hole threshold}
\label{subsec: instantons}

Let us now turn to the sum over D-brane instanton sectors in $Z_N^{\rm grav}$, with $Z_N$ normalized so that the vacuum contributes as $1$.

A stack of $\ell$-coincident D-brane instantons on a gravitational saddle $M$ contributes to $Z_N$ as $e^{-N^2 I_M} e^{-F_M^{(\ell)}}$, where the background-subtracted free energy $F_M^{(\ell)}$ has the large $N$ expansion
\be \label{eq: brane free energy}
F_M^{(\ell)} = \sum_{h=0}^\infty N^{2-2h} F_h^M(\ell/N).
\ee
A genus $h$ worldsheet with $b$ boundaries on the branes comes with $g_s^{2h-2+b}$, and each boundary carries a Chan-Paton trace that contributes a factor of $\ell$. With $g_s \sim 1/N$, we have
\be
g_s^{2h-2+b} \, \ell^b \sim N^{2-2h} \left( \frac{\ell}{N} \right)^b,
\ee
where $\ell/N$ is proportional to the 't Hooft coupling of the $U(\ell)$ worldvolume theory on the D-brane instantons. $F_h^M(\ell/N)$ collects genus $h$ worldsheets with any number of boundaries. At genus zero,
\be \label{eq: genus zero}
F_0^M(\ell/N) = \sum_{b=1}^\infty \left( \frac{\ell}{N} \right)^b F_{0,b}^M.
\ee
The sum starts at $b=1$ because we've normalized each $e^{-F_M^{(\ell)}}$ by the sphere contribution $e^{-N^2 I_M}$ at $b=0$. The disk $b=1$ gives the action
\be
A_M =  F_{0,1}^M
\ee
of a single D-brane instanton, and the annulus $b=2$ between pairs of branes gives the one-loop term
\be
\gamma_M =  F_{0,2}^M.
\ee
Around the gravitational saddle $M$, we thus have
\be
N^2 F_0^M(\ell/N) = \ell N A_M + \ell^2 \gamma_M + F_{0,3}^M \, \frac{\ell^3}{N} + \cdots.
\ee
At fixed $\ell$, the terms with $b \geq 3$ and $h \geq 1$ are suppressed at large $N$ and the $\ell$-instanton sector reduces to
\be
e^{-F_M^{(\ell)}} = e^{-\ell N A_M - \ell^2 \gamma_M } \left( 1 + O\!\left( 1/N \right) \right).
\ee
(This fails in the 't Hooft limit of the D-brane instantons, where $\ell/N$ is held fixed and finite and $N^2 F_0^M(\ell/N)$ does not truncate at the annulus. There, every $b$ contributes at order $N^2$, while the terms with $h \geq 1$ remain suppressed.)

What happens when we sum over $\ell$? It will be instructive for the moment to treat the instanton number $\ell$ as if it were a continuous variable. As a function of $\ell$, the genus zero exponent is stationary at $\ell_*$ with
\be
\big(F_0^{M}\big)'(\ell_*/N) = 0.
\ee
We see that $\ell_*$ will be of order $N$ since $F_0^M$ depends on $\ell$ only via $\ell/N$. Evaluating the contribution $e^{-N^2 I_M} e^{-F_M^{(\ell)}}$ on $\ell_*$, we find at leading order $e^{-N^2 I_{M^*}}$ with
\be \label{eq: backreacted action}
I_{M^*} = I_M + F_0^M(\ell_*/N).
\ee
Our interpretation of $I_{M^*}$ is that it is the on-shell action of the gravitational saddle $M^*$ that results from the backreaction of order $N$ D-brane instantons on saddle $M$. (Whether this new saddle contributes to $Z_N$ at large positive $N$ is a separate question that we will discuss later.) Because $\ell_*/N$ is of order $1$, every term in the open string expansion \eqref{eq: genus zero} at genus zero contributes to $I_{M^*}$. By contrast, the higher genus free energies $F_{h \geq 1}$ are suppressed by $N^{-2h}$ so these do not contribute to $I_{M^*}$.

For the remainder of this section and in Section \ref{sec: index}, we focus on the case in which only the disk and annulus contribute, i.e.\ $F_{0,b \geq 3}^M = 0$. We return to considering general $F_{0,b}^M$ in Section \ref{sec: general}. In this case, we have $\ell_* = -\frac{NA_M}{2\gamma_M}$ and thus the action of saddle $M^*$ relative to that of $M$ is
\be
I_{M^*} - I_M = -\frac{A_M^2}{4\gamma_M}.
\ee
The sign of $\re(I_{M^*} - I_M)$, which depends on the phases of $A_M$ and $\gamma_M$, determines whether saddle $M^*$ dominates over $M$, or vice versa. 

Motivated by the $\frac{1}{16}$-BPS index of Section \ref{sec: index}, we propose that, in the black hole phase (i.e.\ when a black hole saddle contributes and dominates), there exists a family of D-brane instantons with
\be
\re\gamma_M<0.
\ee
This is the situation in which the one-loop factors $e^{-\ell^2 \gamma_M}$ grow with $\ell$ and thus the interactions between the D-brane instantons are enhanced. As we will see shortly, it is also the condition that requires us to accelerate to the level-2 Borel plane. (This condition generalizes to \eqref{eq: general criterion} for general $F_{0,b}^M$.) The converse is not true in general: a family with $\re\gamma_M<0$, whose D-brane instanton sum diverges, may also exist outside the black hole phase.

So far we have not specified the saddle $M$. In the following, we write
\be
I = I_{M^*} - I_M
\ee
for the action of the backreacted saddle $M^*$ relative to that of the starting background $M$. We now discuss in turn the two natural choices for $M$ and $M^*$.

\subsubsection*{From thermal AdS to the black hole}

First consider the case where saddle $M$ is thermal AdS, i.e.\ $M = {\rm AdS}$, which is subleading in the black hole phase.

Thermal AdS has vanishing action $I_{\rm AdS} = 0$ according to our normalization of $Z_N$. Dropping the label $M$ on various quantities on thermal AdS, the action $I \equiv I_{M^*}$ of the backreacted saddle $M^* = {\rm AdS}^*$ is
\be
I = -\frac{A^2}{4 \gamma}.
\ee
This is the situation relevant for the $\frac{1}{16}$-BPS index of Section \ref{sec: index} where the D-brane instantons are giant gravitons. The worldvolume indices of giant gravitons will be independent of $N$ so the terms with $b \geq 3$ turn out to vanish exactly: $F_{0,b \geq 3} = 0$. Moreover, the backreacted action $I$ turn out to be precisely the $\frac{1}{16}$-BPS black hole free energy $I = I_{\rm BH} = F_{\rm BH}/N^2$. 

In the gravitational path integral, the black hole dominates over the vacuum when it contributes and $\re I < 0$. If the D-brane instantons had pure imaginary action $A$, we would have
\be
\re I = \frac{|A|^2 \re\gamma}{4|\gamma|^2},
\ee
in which case $\re I < 0$ would coincide with $\re\gamma < 0$. For general complex $A$, the two conditions differ. Note also that $\re I < 0$ does not by itself indicate whether the black hole saddle contributes. In the $\frac{1}{16}$-BPS index, for example, there are large regions in the space of chemical potentials where $\re I < 0$ holds but the black hole does not contribute.

\subsubsection*{From the black hole to thermal AdS}

Now consider the case where saddle $M$ is an AdS black hole, i.e.\ $M = {\rm BH}$, which is dominant in the black hole phase. We denote the black hole action relative to that of thermal AdS as $I_{\rm BH}$, and we write $\tilde A$ and $\tilde \gamma$ for the action and one-loop term of a set of D-brane instantons on the black hole background. Here, the contribution of the $\ell$-instanton sector in $Z_N^{\rm grav}$ takes the form
\be
e^{-N^2 I_{\rm BH}} e^{-\ell N \tilde{A} - \ell^2 \tilde\gamma}
\ee
up to the annulus order.

Suppose our set of D-brane instantons is such that their backreaction, in the sense that we've been describing, returns us to thermal AdS, i.e.\ $M^* = {\rm BH}^* = {\rm AdS}$. In such a case, the stationary point of the sum in $\ell$ must have the relative action $I = -I_{\rm BH}$ so that the total leading contribution (i.e.\ the absolute action) is consistent with our normalization $I_{\rm AdS} = 0$ in $Z_N$: $e^{-N^2 I_{\rm BH}} e^{-N^2 I} = 1$. That would now require that the following be true of the backreacted action $I$:
\be
I = -I_{\rm BH} = - \frac{\tilde{A}^2}{4 \tilde\gamma}.
\ee
What are the conditions on the data $\tilde{A}, \tilde\gamma$ of the above D-brane instantons that would make the black hole dominate? If the D-brane instantons on the black hole had real action $\tilde A$, we would have
\be \label{eq: BH real}
\re I = -\frac{|\tilde A|^2 \re\tilde\gamma}{4|\tilde\gamma|^2}.
\ee
Then black hole dominance $\re I = -\re I_{\rm BH} > 0$ coincides with the condition that the open string one-loop terms are enhanced with $\ell$: $\re\tilde\gamma<0$. (We would also want $\re \tilde A >0$ so that the instantons are suppressed, though the sign of real $\tilde A$ doesn't affect \eqref{eq: BH real}.)

In the $\frac{1}{16}$-BPS index of Section \ref{sec: index}, we make this relation between the two starting backgrounds more explicit. There, we find that the sum over D3 giant gravitons on thermal AdS, whose collective contribution gives the leading black hole action $I_{\rm BH}$, can be re-expanded around the black hole saddle using the tools of Section \ref{sec: higher}. This allows us to read off the actions $\tilde A$ and one-loop terms $\tilde\gamma$ of the D-brane instantons on the black hole background, from the actions $A$ and one-loop terms $\gamma$ of the D-brane instantons on thermal AdS. The actions $\tilde A$ agree with those of wrapped D3-brane solutions on the black hole background found in Appendix D of \cite{Aharony:2021zkr}. The one-loop terms $\tilde\gamma$ are predictions for the large $\ell$ free energies of those branes.

More generally, using the higher resummation tools of Section \ref{sec: higher}, we show in Section \ref{sec: index} and \ref{sec: general} that the D-brane instanton data $(\tilde{A},\tilde\gamma)$ on the black hole background is related to that $(A,\gamma)$ on thermal AdS as follows, for the case $F_{0,b\geq 3} = 0$: 
\ieg \label{eq: dual data}
\tilde A = \frac{i\pi A}{\gamma}, \qquad \tilde\gamma = \frac{\pi^2}{\gamma}, \qquad I_{\rm BH} = -\frac{A^2}{4\gamma} = \frac{\tilde A^2}{4\tilde\gamma}.
\feg
Since we have that
\be
\re\tilde\gamma = \frac{\pi^2 \re\gamma}{|\gamma|^2},
\ee
the condition $\re\tilde\gamma < 0$ on the black hole background is equivalent to $\re\gamma < 0$ on thermal AdS, as a function of the temperature, chemical potentials, and 't Hooft coupling. Thus the condition $\re\gamma_M < 0$ of our proposal is compatible with both backgrounds.
\be
\re I_{\rm BH} = - \frac{1}{4} \re \! \left( A^2/\gamma \right) = \frac{1}{4} \re \! \left( \tilde{A}^2/ \tilde\gamma \right),
\ee
whether one saddle dominates over another is determined by the phase of $A^2/\gamma$. The black hole dominates over thermal AdS when $\re ( A^2/\gamma) >0$, or equivalently when $\re ( \tilde{A}^2/\tilde\gamma) <0$. For real negative $\gamma$ (and hence real negative $\tilde\gamma$), this means in the black hole phase that we should find $A$ that is mostly imaginary while finding $\tilde A$ that is mostly real. All in all, the above considerations suggest that we can view the two D-brane instanton expansions as being related via a change of the starting background $M$.

In what follows, we return to the instanton data $(A,\gamma)$ around thermal AdS and write $I = I_{\rm BH}$ for the action of the backreacted black hole relative to that of thermal AdS.

\subsection{Obstruction to summability and a precision bound}
\label{subsec: obstruction}

Let us now study what the criterion $\re\gamma < 0$ would mean for the Borel transform $B_1(\zeta_1)$ whose singularities encode the D-brane instanton sum. With $F_{0,b \geq 3} = 0$, the $\ell$-instanton sector is $e^{-F^{(\ell)}} = e^{-\ell N A - \ell^2 \gamma }$ and so the Stokes data of \eqref{eq: multi disc} are given by $c_0^{(\ell)} = e^{-\ell^2\gamma}$ and $c_{n \geq 1}^{(\ell)} = 0$. The discontinuity across $\theta_1 = \arg A$ is\footnote{Additional coefficients $\sigma_\ell$ in the $\ell$-instanton terms of the discontinuity are allowed and do not change the discussion below as long as $\log|\sigma_\ell|$ grows more slowly than $\ell^2$. \label{foot: instanton coeff}}
\be \label{eq: one-loop disc}
\Disc_{\theta_1} B_1(\zeta_1) = \sum_{\ell=1}^\infty e^{-\ell^2\gamma} \, \Theta(|\zeta_1| - \ell|A|).
\ee
Suppose we take $\zeta_1$ on the ray such that
\be
L |A| < |\zeta_1| < (L+1) |A|
\ee
for $L$ a positive integer. Then we have the finite sum
\be \label{eq: disc B1 finite bound}
\Disc_{\theta_1} B_1(\zeta_1) = \sum_{\ell=1}^L e^{-\ell^2\gamma}
\ee
where the dependence on $\zeta_1$ now enters through $L = \lfloor |\zeta_1|/|A| \rfloor$. For $\re \gamma < 0$, this sum is dominated by the $L$-th instanton. This means that the discontinuity grows as
\be
\log|\Disc_{\theta_1} B_1(\zeta_1)| = \frac{|\re\gamma|}{|A|^2}\,|\zeta_1|^2 + O(|\zeta_1|),
\ee
which is the growth \eqref{eq: B1 growth N} with $\rho = \frac{|A|^2}{4|\re\gamma|}$. Thus we find that a line of D-brane instanton singularities on the Borel $\zeta_1$-plane with $\re\gamma <0$ results in the divergence of the level-1 Laplace integral. That is, the level-1 Laplace integral fails to reproduce the original Stokes discontinuity $\sum_{\ell=1}^\infty e^{-\ell N A - \ell^2 \gamma}$.

\subsubsection*{Obstruction to summability}

We now consider the acceleration of the discontinuity $\Disc_{\theta_1} B_1(\zeta_1)$ to the critical level $k'=2$. By \eqref{eq: acc disk} with $\theta_1 = \arg A$, the acceleration integral converges in the disk
\be \label{eq: one-loop disk}
\left| \zeta_2 - \frac{A^2}{8|\re\gamma|} \right| < \frac{|A|^2}{8|\re\gamma|}.
\ee
The circle at the boundary of this disk includes the origin as well as the stationary point $I = -\frac{A^2}{4\gamma}$ from earlier. Applying acceleration termwise as \eqref{eq: acc erfc}, we have the following discontinuity on the $\zeta_2$-plane:
\be \label{eq: one-loop acc}
\Disc_{\theta_2} B_2(\zeta_2) = \sum_{\ell=1}^\infty e^{-\ell^2\gamma} \, \erfc\!\left( \frac{\ell A}{2\sqrt{\zeta_2}} \right).
\ee
One can find using, say, the asymptotic expansion of $\erfc$ at large argument that this sum also converges in the same disk \eqref{eq: one-loop disk}.\footnote{For $F_{0,b \geq 3} = 0$, the acceleration integral and the termwise-accelerated sum converge in the same region in $\zeta_2$. This will not necessarily be the case in the more general setup of Section \ref{sec: general}.}

Making sense of the level-2 Laplace integral (and thus resummation) requires that the discontinuity $\Disc_{\theta_2} B_2(\zeta_2)$ be analytically continuable along the full ray at angle $\theta_2$ on the $\zeta_2$-plane. So let us ask whether $\Disc_{\theta_2} B_2(\zeta_2)$ can be continued beyond its convergence region. To study its analytic properties, it is instructive to take the derivative of the discontinuity with respect to $\zeta_2$:
\be \label{eq: derivative disc}
\frac{\partial}{\partial\zeta_2} \Disc_{\theta_2} B_2(\zeta_2) = \frac{A}{2\sqrt{\pi}\,\zeta_2^{3/2}} \sum_{\ell=1}^\infty \ell \, \mathsf{q}^{\ell^2}
\ee
where we define
\be
\mathsf{q} = \exp\left[ -\left( \frac{A^2}{4\zeta_2} + \gamma \right) \right].
\ee
This series converges in the region $|\mathsf{q}| < 1$, which is in fact the disk \eqref{eq: one-loop disk}.

The Fabry gap theorem \cite{Puchta:2002:Fabry} says that a power series
\be
F(\mathsf{q}) = \sum_\ell a_\ell \, \mathsf{q}^{n_\ell}
\ee
with radius of convergence $1$ and $n_\ell/\ell \to \infty$ cannot be analytically continued across any point of the circle $|\mathsf{q}| = 1$, as every point of the circle is a singular point of $F(\mathsf{q})$. For our $\mathsf{q}$-series \eqref{eq: derivative disc}, we have $n_\ell = \ell^2$ and thus $n_\ell / \ell = \ell \to \infty$. Therefore, the circle
\be \label{eq: one-loop circle}
\left| \zeta_2 - \frac{A^2}{8|\re\gamma|} \right| = \frac{|A|^2}{8|\re\gamma|}
\ee
bounding the convergence disk \eqref{eq: one-loop disk} is a natural boundary of $\pa_{\zeta_2} \Disc_{\theta_2} B_2(\zeta_2)$. This circle is then also a natural boundary of $\Disc_{\theta_2} B_2(\zeta_2)$. To summarize, the level-2 Borel discontinuity $\Disc_{\theta_2} B_2(\zeta_2)$ for $\re\gamma <0$ is defined only in the bounded interior of the disk \eqref{eq: one-loop disk} in the $\zeta_2$-plane, and we cannot continue it outside this disk. \textit{We thus find an analytic obstruction to summability: the level-2 Laplace integral of $\Disc_{\theta_2} B_2(\zeta_2)$ cannot be performed along the ray at angle $\theta_2$ or along any other ray going out to infinity.}

While we've been primarily considering the case $\re\gamma <0$ where accelerating to the $\zeta_2$-plane was necessary, a natural boundary is actually present for either sign of $\re\gamma$. In the case $\re\gamma > 0$, there is no obstruction to performing the level-1 Laplace integral of $\Disc_{\theta_1} B_1(\zeta_1)$. Nevertheless, the identity \eqref{eq: acc stokes}\footnote{This holds as an identity when both sides are well-defined. If only the RHS is defined, the RHS serves as a definition for the LHS.}
\be \label{eq: acc stokes 2}
\cL_1^{\theta_1} \Disc_{\theta_1} B_1 = \cL_2^{\theta_2} \Disc_{\theta_2} B_2
\ee
tells us that we are free to accelerate $\Disc_{\theta_1} B_1$. Accelerating gives \eqref{eq: one-loop acc} with $\re\gamma >0$, and Fabry again tells us that $\Disc_{\theta_2} B_2$ has a natural boundary at $|\mathsf{q}|=1$. The difference, however, is that now the convergence region $|\mathsf{q}|<1$ of $\Disc_{\theta_2} B_2$ corresponds in the $\zeta_2$-plane to
\be \label{eq: one-loop exterior}
\left| \zeta_2 + \frac{A^2}{8 |\re\gamma|} \right| > \frac{|A|^2}{8 |\re\gamma|},
\ee
i.e.\ the unbounded exterior of a circle on the $\zeta_2$-plane. The natural boundary circle again includes the origin and $I = -\frac{A^2}{4 \gamma}$. The unbounded region \eqref{eq: one-loop exterior} allows a ray to go out to infinity at angle $\theta_2$ (as well as every angle within $\frac{\pi}{2}$ of $\theta_2$).\footnote{The ray at angle $\theta_2$ points directly away from the center of the circle since $\arg(-A^2) = \theta_2 + \pi$.} Thus, in the case $\re\gamma >0$, there is no analytic obstruction to performing the level-2 Laplace integral of $\Disc_{\theta_2} B_2(\zeta_2)$ along a ray at angle $\theta_2$, as should be if the identity \eqref{eq: acc stokes 2} holds.

\subsubsection*{Sum versus integral}

Let us pause to comment on what happens if we instead treat $\ell$ as a continuous variable as we did in Section \ref{subsec: instantons} to find $I$. Doing so in the sum \eqref{eq: one-loop acc} for the level-2 Borel discontinuity $\Disc_{\theta_2} B_2(\zeta_2)$, we get\footnote{integrating from $0$}
\be \label{eq: sum integral}
\int_0^\infty d\ell \, e^{-\ell^2\gamma} \, \erfc\!\left( \frac{\ell A}{2\sqrt{\zeta_2}} \right) = \frac{1}{\sqrt{\pi\gamma}} \arctan\!\left( \frac{2\sqrt{\gamma\zeta_2}}{A} \right).
\ee
This has no natural boundary and can be analytically continued beyond the region in which the integral \eqref{eq: sum integral} converges. It also has a logarithmic singularity at $\zeta_2 = I = -\frac{A^2}{4 \gamma}$, where locally
\be \label{eq: sum integral 2}
\eqref{eq: sum integral} \, = \, \frac{-1}{2\sqrt{-\pi \gamma}} \log(\zeta_2 - I) \, + \, \text{regular}.
\ee
The level-2 Stokes discontinuity that would have resulted from crossing this singularity is
\be
i\sqrt{\frac{\pi}{-\gamma}} \, e^{-N^2 I}.
\ee
According to our interpretation of $I$ near \eqref{eq: backreacted action}, this takes the form of a contribution from a gravitational saddle that would result from backreacting the D-brane instantons we started with. We see that the integral approximation produces an ordinary isolated logarithmic singularity and thus the natural boundary was a consequence of the discreteness of $\ell$.

\subsubsection*{A precision bound}

We now return to the level-2 Borel discontinuity $\Disc_{\theta_2} B_2$ with $\re\gamma<0$, which is defined only within the disk \eqref{eq: one-loop disk} bounded by a natural boundary. While $\Disc_{\theta_2} B_2$ admits no Laplace ray extending to infinity, we can nevertheless ask what the best we can do is. That is, despite our inability to perform the resummation, what is the best precision we can achieve using the level-2 Borel data at hand?

Since $Z_N$ is dominated by the black hole above the black hole threshold, we would start in a physical computation from the black hole saddle, compute its perturbative large $N$ expansion, and attempt to apply resummation methods. The D-brane $\ell$-instanton sector on the black hole background contributes up to the annulus order as
\be \label{eq: BH sectors}
e^{-N^2 I_{\rm BH}} \, e^{-\ell N\tilde A - \ell^2\tilde\gamma}
\ee
where $\re \tilde\gamma < 0$ and where $(\tilde A,\tilde\gamma)$ is related to $(A,\gamma)$ as \eqref{eq: dual data} as we derive from the level-2 Borel data in Section \ref{subsec: backgrounds}. Let $\tilde\zeta_1$ and $\tilde\zeta_2$ be the level-1 and level-2 Borel variables for the large $N$ expansion around the black hole saddle. The D-brane instanton sectors \eqref{eq: BH sectors} would yield a level-1 Borel discontinuity $\Disc_{\tilde{\theta}_1} \tilde{B}_1(\tilde\zeta_1)$ containing terms of the form \eqref{eq: one-loop disc} with $\zeta_1, A, \gamma$ replaced by $\tilde\zeta_1, \tilde A, \tilde\gamma$ where $\tilde\theta_1 = \arg \tilde A$. The level-2 Borel discontinuity $\Disc_{\tilde\theta_2} \tilde{B}_2(\tilde\zeta_2)$ is in turn given by the acceleration of $\Disc_{\tilde{\theta}_1} \tilde{B}_1(\tilde\zeta_1)$ where $\tilde\theta_2 = 2 \tilde\theta_1$. It is again defined only within the disk
\be
\bigg| \tilde\zeta_2 - \frac{\tilde A^2}{8|\re\tilde\gamma|} \bigg| < \frac{|\tilde A|^2}{8|\re\tilde\gamma|},
\ee
bounded by a natural boundary through the black hole at $\tilde\zeta_2=0$ and thermal AdS at $\tilde\zeta_2=-I_{\rm BH}$.

We would like to represent the expansions around the black hole and thermal AdS saddles on the same level-2 Borel plane by writing $\zeta_2 = I_{\rm BH} + \tilde\zeta_2$. At the level of the level-2 Laplace kernel, this only amounts to a change of variables $e^{-N^2\zeta_2} = e^{-N^2 I_{\rm BH}} e^{-N^2\tilde\zeta_2}$ where the points of the $\zeta_2$-plane represent absolute actions in our normalization $I_{\rm AdS}=0$. The non-trivial physical step underlying this change of variables is that the level-2 Borel data around the two saddles should be compatible, i.e. that $\Disc_{\tilde\theta_2} \tilde{B}_2(\tilde\zeta_2)$ be a part of the same analytic structure on the $\zeta_2$-plane as $\Disc_{\theta_2} B_2(\zeta_2)$. We can understand this as the level-2 analog of the familiar resurgence statement that the Borel transform around one saddle encodes the expansion around another saddle. As a check, \eqref{eq: dual data} implies that the natural boundary of $\Disc_{\tilde\theta_2} \tilde{B}_2$ is also the circle \eqref{eq: one-loop circle} on the $\zeta_2$-plane.

On the $\zeta_2$-plane, a level-2 resummation around the black hole saddle in the black hole phase would thus correspond to starting the level-2 Laplace integral at $\zeta_2 = I_{\rm BH}$ (or equivalently $\tilde\zeta_2 = 0$) on the natural boundary, as opposed to starting at $\zeta_2 = 0$. Since the disk is bounded, we need to truncate the level-2 Laplace integral starting at $I_{\rm BH}$ at some point $\zeta_2^*$ before reaching the natural boundary, where the part of the integral beyond $\zeta_2^*$ corresponds to an ``error'' that is undetermined by the Borel data. This remainder is of order $e^{-N^2\re\zeta_2^*}$ due to the Laplace kernel $e^{-N^2\zeta_2}$. We can achieve the greatest suppression for the remainder term by taking $\zeta_2^*$ to approach the rightmost point of the disk \eqref{eq: one-loop disk}, whose real part $\hat I$ is
\be \label{eq: one-loop Shat}
\hat I = \frac{(\re A)^2}{4|\re\gamma|} = \frac{|\tilde\gamma|^2\,(\im(\tilde A/\tilde\gamma))^2}{4|\re\tilde\gamma|} \geq 0.
\ee
For pure imaginary $A$ on thermal AdS or for pure real $\tilde A/\tilde\gamma$ on the black hole background, the rightmost point is the origin of the $\zeta_2$-plane. Relative to the black hole contribution, the remainder term is of order $e^{-N^2(\hat I-\re I_{\rm BH})}$ where
\be \label{eq: relative Shat}
\hat I - \re I_{\rm BH} = \frac{|\gamma|^2\,(\im(A/\gamma))^2}{4|\re\gamma|} = \frac{(\re\tilde A)^2}{4|\re\tilde\gamma|}.
\ee

\paragraph{Proposal}

Let $Z_N$ be normalized so that the vacuum contributes as $1$, and suppose that we are in the black hole phase, i.e.\ $Z_N \sim e^{-N^2 I_{\rm BH}}$ with $\re I_{\rm BH}<0$. Suppose further that the gravitational transseries contains D-brane instanton sectors whose level-2 Borel discontinuities around both saddles are confined to a bounded region $\cR$ of the $\zeta_2$-plane by a natural boundary through thermal AdS at $\zeta_2=0$ and the black hole at $\zeta_2=I_{\rm BH}$. Define
\be \label{eq: geometric Shat}
\hat I = \max_{\zeta_2\in\overline{\cR}} \re\zeta_2
\ee
where the closure $\overline{\cR}$ includes the natural boundary.

We propose that a resummation $\overline{Z_N^{\rm grav}}$ of the gravitational transseries $Z_N^{\rm grav}$ can determine $Z_N$ only up to
\be \label{eq: precision bound}
|Z_N-\overline{Z_N^{\rm grav}}| \, \gtrsim \, O \big( e^{-N^2\hat I} \big).
\ee
Here, the inequality $\gtrsim$ compares only the coefficients of $N^2$ in the exponent, so any subleading factors such as $e^{-N}$ or powers of $1/N$ are left unconstrained. Relative to the black hole, this corresponds to the normalized error
\be \label{eq: precision bound norm}
\frac{|Z_N-\overline{Z_N^{\rm grav}}|}{|Z_N|} \gtrsim O \big(e^{-N^2(\hat I-\re I_{\rm BH})}\big)
\ee
which is exponentially small.

For $F_{0,b\geq3}=0$, a natural boundary enclosing a bounded $\cR$ arises precisely when $\re\gamma_M<0$, with $\cR$ the disk \eqref{eq: one-loop disk} and $\hat I$ given by \eqref{eq: one-loop Shat}.

\vspace{2em}

Let us conclude by comparing the precision scale $e^{-N^2\hat{I}}$ with the optimal truncation of the D-brane instanton sum in the black hole background with $F_{0,b\geq3}=0$. For $\re\tilde A>0$ and $\re\tilde\gamma<0$, the magnitudes
\be
\left| e^{-N^2 I_{\rm BH}}e^{-\ell N\tilde A-\ell^2\tilde\gamma} \right| = e^{-N^2\re I_{\rm BH}-\ell N\re\tilde A+\ell^2|\re\tilde\gamma|}
\ee
of the D-brane instanton terms around the black hole decrease until $\ell_{\rm optimal}= \frac{N\re\tilde A}{2|\re\tilde\gamma|}$ and grow beyond it. The term at $\ell=\ell_{\rm optimal}$ evaluates to the magnitude
\be
\exp\!\left[ -N^2\left( \re I_{\rm BH} + \frac{(\re\tilde A)^2}{4|\re\tilde\gamma|} \right) \right] = e^{-N^2\hat I}.
\ee
We find that the proposed scale of the precision bound agrees with the scale of optimal truncation of the D-brane instanton sum in the black hole background.\footnote{A relevant work \cite{Pioline:2009ia} considered a D-instanton sum with a gaussian annulus growth and found an $e^{-1/g_s^2}$ ambiguity in the optimal truncation.}

We can contrast the situation here with the usual optimal truncation of the large $N$ perturbative expansion whose coefficients grow as $(2h)!$. There, the optimal truncation scale $e^{-N A}$ is the size of a D-brane instanton contribution, and Borel resummation can improve on this precision via the analytic continuation in $\zeta_1$ of the level-1 Borel transform beyond the initial disk of convergence. Here, the natural boundary obstructs the continuation in $\zeta_2$ for the level-2 Borel transform. Our proposal is that this obstruction makes the scale $e^{-N^2\hat I}$ a limit on the precision that is attainable from the gravitational transseries, rather than being the size of a next isolated saddle contribution. It would nevertheless be important to understand the physical origin of the dense set of singularities that form the natural boundary, and we will further comment on these dense singularities in Section \ref{subsec: dense}.

So far, our discussion assumed that only the disk and annulus contribute at genus zero. But we saw in Section \ref{subsec: instantons} that every term in $F_0^M(\ell/N)$ can contribute at order $N^2$ in the exponent when $\ell$ is of order $N$. The contributions $F_{0,b\geq3}$ can therefore change $\cR$ and $\hat I$. Our proposal \eqref{eq: geometric Shat}--\eqref{eq: precision bound norm} still applies in that case provided that a natural boundary continues to confine the level-2 Borel data to a bounded region $\cR$. We will return to this point in Section \ref{sec: general} after studying the index for which $F_{0,b\geq 3}=0$.

\section{Black holes and the $\frac{1}{16}$-BPS index}
\label{sec: index}

In this section, we study the precision bound for the gravitational transseries representation of the $\frac{1}{16}$-BPS index of $U(N)$ $\cN=4$ super Yang-Mills theory, considered above the black hole threshold.

\subsection{Preliminaries}
\label{sec: preliminaries}

Let us now consider an example in which we know the black hole and the D-brane instantons: the $\frac{1}{16}$-BPS index of 4d $U(N)$ $\cN = 4$ super Yang-Mills theory on $S^3$.

Let $R_{1,2,3}$ be the Cartan charges of $SO(6)_R$, $J_{1,2}$ the two angular momenta on $S^3$, and $\cQ$ a supercharge with $\{ \cQ, \cQ^\dag \} = H - J_1 - J_2 - R_1 - R_2 - R_3$. The index is defined as \cite{Kinney:2005ej,Romelsberger:2005eg}
\be
Z_N = \Tr_{\cH_{S^3}} \left( (-1)^F e^{-\beta \{ \cQ, \cQ^\dag \}} p^{J_1} q^{J_2} x^{R_1} y^{R_2} z^{R_3} \right)
\ee
with the constraint $xyz=pq$ so that the charges form four independent combinations that commute with $\cQ, \cQ^\dag$. The index counts, with signs
\be
(-1)^F = e^{2 \pi i J_{1,2}} = e^{2 \pi i R_{1,2,3}},
\ee
the states with $\{ \cQ, \cQ^\dag \} = 0$. The trace is normalized so that the vacuum contributes as $1$. We write
\be
( x,y,z,p,q ) = ( e^{2 \pi i \Delta_1}, e^{2 \pi i \Delta_2}, e^{2 \pi i \Delta_3}, e^{2 \pi i \sigma},  e^{2 \pi i \tau} )
\ee
constrained as
\be
\sigma + \tau - \Delta_1 - \Delta_2 - \Delta_3 = \pm 1
\ee
as required for the interpretation of $Z_N$ in terms of supersymmetric Euclidean black holes. In a region of complexified chemical potentials where the black hole dominates, it was found \cite{Cabo-Bizet:2018ehj,Choi:2018hmj,Benini:2018ywd,Benini:2018mlo} that the index behaves as $Z_N \sim e^{-F_{\rm BH}}$ where 
\be
F_{\rm BH} = i \pi N^2 \frac{\Delta_1 \Delta_2 \Delta_3}{\sigma \tau},
\ee
which yields an entropy consistent with that of $\frac{1}{16}$-BPS black holes in AdS$_5$ \cite{Gutowski:2004ez,Gutowski:2004yv,Kunduri:2006ek}. 

The D-brane instantons in this context are giant gravitons, which are D3-branes wrapping the three maximal $S^3$'s in $S^5$ that are each fixed under the $U(1)_{R_1}$, $U(1)_{R_2}$, and $U(1)_{R_3}$ isometry actions \cite{McGreevy:2000cw}. With $k_1$, $k_2$, $k_3$ giants on the three $S^3$'s, the $\frac{1}{16}$-BPS index at finite $N$ admits the giant graviton expansion \cite{Imamura:2021ytr,Gaiotto:2021xce} (see also \cite{Arai:2020qaj,Murthy:2022ien,Lee:2022vig,Imamura:2022aua,Choi:2022ovw,Liu:2022olj,Eniceicu:2023uvd,Beccaria:2023hip,Lee:2023iil,Eleftheriou:2023jxr,Beccaria:2024vfx,Kim:2024ucf,Gautason:2024nru,Chen:2024cvf,Ezroura:2024wmp,Chen:2024erz,Lee:2024hef,Deddo:2025lfm})
\be \label{eq: GGE}
Z_N = Z_\infty \sum_{k_1, k_2, k_3 = 0}^\infty x^{k_1 N} y^{k_2 N} z^{k_3 N} \, \hat{Z}_{(k_1, k_2, k_3)}.
\ee
$Z_\infty$ is the index of bulk Kaluza-Klein modes, $x^{k_1 N} y^{k_2 N} z^{k_3 N}$ are the classical contributions of the giant graviton instantons, and $\hat{Z}_{(k_1,k_2,k_3)}$ is the index of the $U(k_1) \times U(k_2) \times U(k_3)$ gauge theory on their worldvolumes, where $\hat Z_{(0,0,0)} = 1$. The worldvolume index at large ranks has a saddle with free energy \cite{Chen:2024erz}
\be \label{eq: worldvolume F}
\hat F_{\mathbf{k}} = -i\pi \frac{\sigma\tau}{\Delta_1\Delta_2\Delta_3} \, (\mathbf{k}\cdot\mathbf{\Delta})^2,
\ee
where $\mathbf{k} = (k_1,k_2,k_3)$ and $\mathbf{k}\cdot\mathbf{\Delta} = k_1\Delta_1 + k_2\Delta_2 + k_3\Delta_3$. We will be interested in the region of chemical potentials where this saddle dominates as $\hat Z_{\mathbf{k}} \sim e^{-\hat F_{\mathbf{k}}}$. For a single stack $(k,0,0)$ of $k$-coincident D3 giants, we have
\be
\hat F_{(k,0,0)} = - i\pi k^2 \frac{\Delta_1 \sigma \tau}{\Delta_2 \Delta_3}.
\ee
This is related to $F_{\rm BH}$, with $N$ replaced by $k$ and with the chemical potentials permuted as
\be
(\Delta_1, \Delta_2, \Delta_3, \sigma, \tau) \, \to \, (-\Delta_1, \sigma, \tau, \Delta_2, \Delta_3).
\ee
The worldvolume index $\hat Z_{(k,0,0)}$ is related to the boundary index $Z_N$ via these permutations due to the orientation of the D3 giants relative to the backreacting D3-branes \cite{Gaiotto:2021xce} (it also involves as different integration cycle over the gauge holonomies \cite{Lee:2022vig,Imamura:2022aua}). We assume that, in the region of chemical potentials under our consideration, this large $\mathbf{k}$ saddle contributes and dominates.

On the level-1 Borel plane, the giants produce singularities at $\zeta_1 = -2\pi i(\mathbf{k}\cdot\mathbf{\Delta})$ that fill a wedge spanned by $-2\pi i\Delta_1$, $-2\pi i\Delta_2$, $-2\pi i\Delta_3$.\footnote{For the superconformal index, the perturbative large $N$ expansion of $Z_N$ on AdS$_5 \times S^5$ is one-loop exact and given by the KK spectrum. These Borel singularities should therefore be understood as in Cheshire cat resurgence \cite{Dunne:2016jsr,Kozcaz:2016wvy}, i.e.\ we assume that, once we deform away from the BPS limit, the perturbative coefficients acquire the growth that produces these singularities. We take the BPS limit after extracting the singularities. Assuming this, we determine $\Disc_{\theta_1}B_1$ directly from the giant graviton expansion.} Let us write
\be
\mathbf{k} = \ell \mathbf{n} = \ell (n_1,n_2,n_3)
\ee
where $\gcd(n_1,n_2,n_3)=1$. The singularities along $\mathbf{n}$ are those of an instanton sum with $F_{0,b\geq3} = 0$ and
\ie \label{eq: index data}
A_{\mathbf{n}} &= -2\pi i \, (\mathbf{n}\cdot\mathbf{\Delta}) \\
\gamma_{\mathbf{n}} &= -i\pi \frac{\sigma\tau}{\Delta_1\Delta_2\Delta_3} \, (\mathbf{n}\cdot\mathbf{\Delta})^2,
\fe
and we are interested in the chemical potential region where $\re\gamma_{\mathbf{n}}<0$ for at least one direction $\mathbf n$.

For every direction $\mathbf{n}$, the action $I = -\frac{A^2}{4\gamma}$ of the backreacted saddle, in the sense that we've been describing, is that of the $\frac{1}{16}$-BPS black hole:
\be \label{eq: index I}
I = -\frac{A_{\mathbf{n}}^2}{4\gamma_{\mathbf{n}}} = i\pi \frac{\Delta_1\Delta_2\Delta_3}{\sigma\tau} = \frac{1}{N^2} F_{\rm BH}.
\ee
This is the action resulting from the stationary point $\ell_* = -\frac{N A_{\mathbf{n}}}{2 \gamma_{\mathbf{n}}}$ of order $N$ in the sum over the number of giants along direction $\mathbf{n}$ \cite{Choi:2022ovw,Beccaria:2023hip,Kim:2024ucf,Chen:2024erz}. Stacks of order $N$ number of giants, in any proportion $\mathbf n$, therefore backreact into the black hole. Furthermore, the natural boundaries \eqref{eq: one-loop circle} for any $\mathbf{n}$ pass through both the vacuum $\zeta_2 = 0$ and the black hole $\zeta_2 = I$. If we had $\re\gamma_{\mathbf{n}} \geq 0$ for all $\mathbf n$, thermal AdS would dominate because the annulus term would not be able to overcome the classical suppression of the giants, i.e.\ $Z_N/Z_\infty = 1 + O(e^{-N})$. A dominant black hole thus requires that $\re\gamma_{\mathbf{n}} < 0$ for some $\mathbf n$.

\subsection{D-brane instantons across backgrounds}
\label{subsec: backgrounds}

Let us now derive the relation \eqref{eq: dual data} by relating the D-brane instantons on thermal AdS to those on the black hole. The discussion here applies to any instanton sum with $F_{0,b\geq 3}=0$. At the end, we apply these results to the $\frac{1}{16}$-BPS black hole and its D-brane instantons. We will generalize these results to $F_{0,b\geq 3} \neq 0$ in Section \ref{sec: general}.

\subsubsection*{D-brane instantons on the black hole background}

Around thermal AdS, we found in \eqref{eq: one-loop acc} that the discontinuity of the level-2 Borel transform takes the form
\be \label{eq: one-loop acc 2}
\Disc_{\theta_2}B_2(\zeta_2) = \sum_{\ell=1}^\infty e^{-\ell^2\gamma} \, \erfc \! \left( \frac{\ell A}{2\sqrt{\zeta_2}} \right).
\ee
A formal termwise application of the level-2 Laplace integral along $\theta_2 = 2\theta_1 = 2\arg A$, i.e.\ the ray through the center of the disk \eqref{eq: one-loop disk}, returns the corresponding instanton sector $e^{-\ell NA - \ell^2\gamma}$ for each $\ell$. This holds only termwise for $\re\gamma<0$ since $\Disc_{\theta_2}B_2 (\zeta_2)$ cannot be continued past the natural boundary.

We would like to expand \eqref{eq: one-loop acc 2} around the point corresponding to the black hole action on the $\zeta_2$-plane. Let us write
\be \label{eq: one-loop q}
\mathsf{q} = e^{-\mathsf{t}} =  \exp \! \left[ - \left( \frac{A^2}{4\zeta_2} + \gamma \right) \right]
\ee
so that the disk \eqref{eq: one-loop disk} corresponds to
\be
\re \mathsf{t} = \re \! \left( \frac{A^2}{4\zeta_2} + \gamma \right) > 0.
\ee
The thermal AdS point at the origin of the $\zeta_2$-plane corresponds to taking $\mathsf{t} \to \infty$ or $\mathsf{q} \to 0$, while the black hole point $\zeta_2 = I = -\frac{A^2}{4 \gamma}$ corresponds to taking $\mathsf{t} \to 0$ or $\mathsf{q} \to 1$. To expand around the black hole point, we need to consider a resummation of the instanton sectors. To do so, it is convenient to take the derivative of $\Disc_{\theta_2}B_2$ as in \eqref{eq: derivative disc} as
\be \label{eq: derivative disc 2}
\frac{\pa}{\pa\zeta_2}\Disc_{\theta_2} B_2(\zeta_2) = \frac{A}{2\sqrt{\pi}\,\zeta_2^{3/2}} \sum_{\ell=1}^\infty \ell\,e^{-\mathsf{t}\ell^2},
\ee
where the exponentials from the D-brane instanton sectors are expanded around $\mathsf{t} = \infty$. Let us write the sum over the exponentials as
\be
\sum_{\ell=1}^\infty \ell\,e^{-\mathsf{t}\ell^2} = \int_0^\infty du\,u\,e^{-\mathsf{t}u^2} \sum_{\ell \in \bZ}\delta(u-\ell).
\ee
The sum on the RHS is periodic under integer shifts in $u$, and its Fourier series representation is $\sum_{\ell \in \bZ} \delta(u - \ell) = \sum_{k \in \bZ} e^{2 \pi i k u}$. Substituting this representation and pairing opposite Fourier modes, we obtain
\be \label{eq: fourier}
\sum_{\ell=1}^\infty \ell\,e^{-\mathsf{t}\ell^2} = \frac{1}{2\mathsf{t}} + 2\sum_{k=1}^\infty \int_0^\infty du\,u\,e^{-\mathsf{t}u^2}\cos(2\pi ku).
\ee
The zeroth Fourier mode $k=0$ gives the first term on the RHS as $\int_0^\infty du \, u \, e^{-\mathsf{t} u^2} = \frac{1}{2\mathsf{t}}$, which amounts to replacing the sum by the integral as in \eqref{eq: sum integral}. Furthermore, we express each higher mode $k\geq1$ as 
\be \label{eq: fourier 2}
\int_0^\infty du \, u \, e^{-\mathsf{t}u^2} \cos(2\pi ku) = \int_0^\infty du \, u \, e^{-\mathsf{t}u^2 \mp 2\pi iku} \pm \frac{i\pi^{3/2}k}{2\mathsf{t}^{3/2}} e^{-\pi^2k^2/\mathsf{t}}
\ee
where we can retain either positive or negative Fourier frequencies. The second term on the RHS comes from a gaussian integral on the full line, formed using that $u e^{-\mathsf{t}u^2}$ is odd.

Let $\Psi_{\pm}(\zeta_2)$ be the $\zeta_2$-primitive of the zeroth and positive or negative Fourier terms of $\pa_{\zeta_2} \Disc_{\theta_2}B_2$ in the decomposition \eqref{eq: fourier 2}, so that its derivative is
\be
\frac{\pa\Psi_\pm}{\pa\zeta_2} = \frac{A}{2\sqrt{\pi}\,\zeta_2^{3/2}} \left( \frac{1}{2\mathsf{t}} + 2\sum_{k=1}^\infty \int_0^\infty du \, u \, e^{-\mathsf{t} u^2 \mp 2\pi iku} \right).
\ee
Then we have
\be
\frac{\pa}{\pa\zeta_2}\Disc_{\theta_2}B_2(\zeta_2) = \frac{\pa\Psi_\pm}{\pa\zeta_2} \pm \frac{i \pi A}{2 \zeta_2^{3/2}} \frac{1}{\mathsf{t}^{3/2}} \sum_{k=1}^\infty k \, e^{-\pi^2k^2/\mathsf{t}}.
\ee
The second term on the RHS is a sum of exponentials expanded around $\mathsf{t} = 0$, i.e.\ the black hole point $\zeta_2 = I$. To expose its physical interpretation, let us write the exponentials in terms of $\tilde\zeta_2 = \zeta_2 - I$ as
\be \label{eq: dual data 2}
\tilde{\mathsf{q}} \equiv e^{-\pi^2/\mathsf{t}} = \exp \! \left[ - \left(\frac{\tilde A^2}{4\tilde\zeta_2} + \tilde\gamma \right) \right], \quad \tilde A = \frac{i\pi A}{\gamma}, \quad \tilde\gamma = \frac{\pi^2}{\gamma}.
\ee
Doing so, we get\footnote{We choose roots so that $\re ( A/\sqrt{\zeta_2} )>0$ and $\re ( \tilde A/\sqrt{\tilde\zeta_2} )>0$.}
\be \label{eq: derivative disc 3}
\frac{\pa}{\pa\zeta_2}\Disc_{\theta_2}B_2(\zeta_2) = \frac{\pa\Psi_\pm}{\pa\zeta_2} \mp i\sqrt{\frac{\pi}{-\gamma}} \, \frac{\tilde A}{2\sqrt{\pi}\,\tilde\zeta_2^{3/2}} \sum_{k=1}^\infty k \, e^{-\pi^2 k^2/\mathsf{t}}.
\ee
We identify the final term as the analog of the quantity \eqref{eq: derivative disc 2} around the black hole point:
\be
\frac{\pa}{\pa \tilde\zeta_2} \Disc_{\tilde\theta_2} \tilde B_2(\tilde\zeta_2) = \frac{\tilde A}{2\sqrt{\pi}\,\tilde\zeta_2^{3/2}} \sum_{k=1}^\infty k \, e^{-\pi^2 k^2/\mathsf{t}},
\ee
where $\tilde\theta_2 = 2 \arg \tilde{A}$. Then we find, up to an integration constant,
\be \label{eq: disc tilde}
\Disc_{\tilde\theta_2}\tilde B_2(\tilde\zeta_2) = \sum_{k=1}^\infty e^{-k^2\tilde\gamma} \, \erfc\!\left( \frac{k\tilde A}{2\sqrt{\tilde\zeta_2}} \right).
\ee
A formal termwise application\footnote{The level-2 Laplace integral here is a formal operation because, for $\re\gamma_M<0$, the sum \eqref{eq: disc tilde} is only defined within a bounded region enclosed by a natural boundary. By contrast, going from the instanton data \eqref{eq: BH instantons} on the black hole background to the discontinuity \eqref{eq: disc tilde} is an analytic operation.} of the level-2 Laplace integral on \eqref{eq: disc tilde} starting at $\tilde\zeta_2 = 0$ would return
\be \label{eq: BH instantons}
e^{-N^2 I} \sum_{k=1}^\infty e^{-k N \tilde A - k^2 \tilde\gamma}.
\ee
For $I = I_{\rm BH}$, i.e.\ for the set of D-brane instantons on thermal AdS whose backreaction, in the sense described in Section \ref{subsec: instantons}, yields the on-shell action of a black hole, we interpret \eqref{eq: BH instantons} as the sum over D-brane instantons on the black hole background. \textit{The formulas \eqref{eq: dual data 2} thus relate the D-brane instanton data $(A,\gamma)$ on thermal AdS to the D-brane instanton data $(\tilde A, \tilde\gamma)$ on the black hole background.} We find that $|\tilde{\mathsf{q}}|<1$ and $|\mathsf{q}|<1$ give the same disk, so $\Disc_{\theta_2} B_2$ and $\Disc_{\tilde\theta_2}\tilde B_2$ share the natural boundary \eqref{eq: one-loop circle}.

From the above considerations, we can relate the level-2 Borel data around the thermal AdS and black hole saddles:
\be \label{eq: disc relation}
\Disc_{\theta_2}B_2(\zeta_2) = \Psi_\pm(\zeta_2) \mp i\sqrt{\frac{\pi}{-\gamma}} \, \Disc_{\tilde\theta_2}\tilde B_2(\tilde\zeta_2),
\ee
where we have absorbed the integration constant into $\Psi_\pm$. The remaining term $\Psi_\pm$ has the singularity $\Psi_\pm(\zeta_2) \sim -\frac{1}{2\sqrt{-\pi\gamma}}\log\tilde\zeta_2$ for either sign as we take $\tilde\zeta_2 = \zeta_2 - I \to 0$ from inside the disk \eqref{eq: one-loop disk}. We have seen this logarithmic singularity previously in \eqref{eq: sum integral 2}: it is the level-2 Borel singularity that, if it were isolated, would produce the Stokes discontinuity $i \sqrt{\frac{\pi}{-\gamma}} e^{-N^2 I}$ corresponding to a black hole saddle.

\subsubsection*{Giants on the black hole}

Let us now apply our results to the $\frac{1}{16}$-BPS black hole and its D-brane instantons. 

Giant gravitons on thermal AdS$_5 \times S^5$ along the direction $\mathbf{n}$ have instanton actions $A_{\mathbf{n}} = -2\pi i \, (\mathbf{n}\cdot\mathbf{\Delta})$ and leading one-loop free energies $\gamma_{\mathbf{n}} = -i\pi \frac{\sigma\tau}{\Delta_1\Delta_2\Delta_3} \, (\mathbf{n}\cdot\mathbf{\Delta})^2$, and the higher terms in the expansion of their free energies vanish: $F_{0,b\geq3} = F_{h\geq1,b} = 0$. Their backreacted action $I = -\frac{A_{\mathbf{n}}^2}{4\gamma_{\mathbf{n}}} = i\pi \frac{\Delta_1\Delta_2\Delta_3}{\sigma\tau}$ is the on-shell action $I_{\rm BH}$ of the $\frac{1}{16}$-BPS black hole, for any $\mathbf{n}$, in the region of chemical potentials where the above expressions for $A_{\mathbf{n}}, \gamma_{\mathbf{n}}$ are valid.

Applying \eqref{eq: dual data 2} to the instanton sum along $\mathbf n$, we find
\be \label{eq: index dual data}
\tilde A_{\mathbf n} = \frac{2 I_{\rm BH}}{\mathbf{n}\cdot\mathbf{\Delta}} = 2\pi i \, \frac{\Delta_1\Delta_2\Delta_3}{\sigma\tau\,(\mathbf{n}\cdot\mathbf{\Delta})}, \qquad \tilde\gamma_{\mathbf n} = \frac{I_{\rm BH}}{(\mathbf{n}\cdot\mathbf{\Delta})^2} = i\pi \, \frac{\Delta_1\Delta_2\Delta_3}{\sigma\tau\,(\mathbf{n}\cdot\mathbf{\Delta})^2}.
\ee
For $\mathbf n = (1,0,0)$, i.e.\ a single stack of $\ell$-coincident giants on thermal AdS$_5$ $\times$ S$^5$ whose worldvolume is fixed under the R-isometry $U(1)_{R_1}$, we find the ``dual'' D-brane instanton data on the black hole background
\be \label{eq: BH giant data}
\tilde A = 2\pi i \frac{\Delta_2\Delta_3}{\sigma\tau}, \qquad \tilde\gamma = i\pi \frac{\Delta_2\Delta_3}{\sigma\tau\Delta_1}.
\ee
The cases $\mathbf n = (0,1,0)$ and $(0,0,1)$ are related to the above by permutations. The action $\tilde A$ and its permutations agree with those of wrapped D3-brane solutions on the black hole background found in Appendix D of \cite{Aharony:2021zkr}.\footnote{$e^{-N \tilde A}|_{\rm here} = e^{i S_{\rm D3}}|_{\rm there}$} Here, we obtained them only from the data of the D3 giant gravitons on thermal AdS. These D3-brane instantons sit at the tip of the Euclidean time cigar and wrap the $S^3$ horizon in the AdS$_5$ components and an $S^1$ in $S^5$. The one-loop terms $\tilde\gamma$ of these wrapped D3-branes have not been considered in the literature, as far as we are aware. The agreement of the actions motivates the prediction that the worldvolume index of $k$-coincident such D3-branes on the Euclidean $\frac{1}{16}$-BPS black hole background has a large $k$ saddle with free energy $k^2\tilde\gamma$. Since we have that $I_{\rm BH} - \frac{\tilde A^2}{4\tilde\gamma} = I_{\rm AdS} = 0$, a stack of order $k \sim N$ such D-brane instantons backreact, in the sense described in Section \ref{subsec: instantons}, into thermal AdS.\footnote{The work \cite{Aharony:2021zkr} also considered other D3-brane instantons at the tip of the cigar on the black hole background, wrapped on an $S^1$ of the $S^3$ horizon and on an $S^3$ in $S^5$. It would be very interesting to determine their annulus free energy $\tilde\gamma$, the gravitational saddle into which they backreact, and their ``dual'' D-brane instanton data on that saddle.}

\subsection{Precision bound for the index}
\label{subsection: index bound}

Let us now determine the precision bound for the index above the black hole threshold. While the specific value of the precision bound for the index is not of physical interest, it is notable that the gravitational transseries $Z_N^{\rm grav}$ representation of this protected quantity admits, at all, an obstruction to recovering the exact answer $Z_N$. 

As we saw in Section \ref{sec: preliminaries}, a dominant black hole requires that we have $\re\gamma_{\mathbf n}<0$ for some direction $\mathbf n$. (The presence of such a direction makes the giant graviton expansion \eqref{eq: GGE} diverge when evaluated at chemical potentials relevant for the black hole phase.) Each such direction yields a discontinuity $\Disc_{\theta_2} B_2$, with $\theta_2 = 2 \arg A_{\mathbf n}$, that is defined only within the disk \eqref{eq: one-loop disk}. Each such disk is bounded by a natural boundary, and this boundary includes singularities in $\zeta_2$ whose locations are the on-shell actions of thermal AdS and the black hole.

The giant graviton expansion \eqref{eq: GGE} receives contributions from all terms in the sum over D-brane instanton sectors at fixed finite chemical potentials.\footnote{This is in contrast with a microcanonical evaluation, in which wall-crossing can make the contributions of an infinite subset of terms vanish \cite{Gaiotto:2021xce,Lee:2022vig,Imamura:2022aua}.} The expansion therefore involves the full D-brane instanton sum along each direction $\mathbf n$, and the directions with $\re\gamma_{\mathbf n}<0$ are subject to the resummation obstruction of Section \ref{subsec: obstruction}. From \eqref{eq: one-loop Shat} and \eqref{eq: index data}, the resulting remainder along $\mathbf n$ is of order $e^{-N^2\hat I_{\mathbf n}}$, with
\be \label{eq: index bound n}
\hat I_{\mathbf n} = \frac{(\re A_{\mathbf n})^2}{4|\re\gamma_{\mathbf n}|} = \frac{(\im(\mathbf n\cdot\mathbf\Delta))^2}{|\re((\mathbf n\cdot\mathbf\Delta)^2/I_{\rm BH})|}.
\ee
The precision bound \eqref{eq: precision bound} for the gravitational transseries representation of the index is set by the largest remainder of order $e^{-N^2 \hat I}$, with
\be \label{eq: index bound}
\hat I = \min_{\mathbf n \,:\, \re\gamma_{\mathbf n}<0} \hat I_{\mathbf n}.
\ee

Anticipating the discussion in Section \ref{sec: generic}, one may ask whether the precision bound for the index above the black hole threshold indicates a term in $Z_N$ of size $e^{-N^2\hat I}$ that is erratic in $N$. The precision bound states that the gravitational transseries $Z_N^{\rm grav}$ determines $Z_N$ only up to a remainder of order $e^{-N^2\hat I}$, but it does not imply that this remainder is erratic. It is natural to interpret the scale $e^{-N^2\hat I}$ as an upper bound on the size of an erratic part, if such a term exists at all. Such a contribution would appear as the connected part of a suitable average of $|Z_N|^2$, which in gravity one expects to come from wormhole saddles. But the index does not receive contributions from geometric wormhole saddles \cite{Iliesiu:2021are}. Thus, either the remainder is not erratic or the index has a variance over $N$ under a suitable average that does not come from a connected supergravity solution. In either case, the index example shows that the precision bound need not imply that a suitable average of $|Z_N|^2$ involves a wormhole contribution.

\subsection{Dense singularities and analytic continuation in $N$}
\label{subsec: dense}

An isolated singularity at $\zeta_2$ on the level-2 Borel plane would give a contribution of order $e^{-N^2 \zeta_2}$. Of the dense set of singularities on the natural boundary, we found so far that two, thermal AdS at $\zeta_2 = 0$ and the black hole at $\zeta_2 = I$, have such an interpretation. What is the interpretation of the other dense set of singularities on the natural boundary on the $\zeta_2$-plane? The natural boundary and its singularities are present for either sign of $\re\gamma$. We can thus address this question in the half-BPS partition function. The level-2 Borel discontinuity $\Disc_{\theta_2} B_2(\zeta_2)$ in this sector has a natural boundary but encounters no resummation obstruction since its giant graviton expansion only allows $\re\gamma>0$.

The half-BPS partition function $Z_N = \prod_{n=1}^N (1-x^n)^{-1}$ can be written in the form
\be \label{eq: half-BPS cont}
\frac{Z_N}{Z_\infty} = \prod_{n=1}^\infty \big( 1 - e^{-\mu(N+n)} \big),
\ee
which is entire in $N$ and thus admits an analytic continuation, where $x = e^{-\mu}$ is the relevant R-charge fugacity. It has the giant graviton expansion \cite{Gaiotto:2021xce}
\be
\frac{Z_N}{Z_\infty} = \sum_{\ell=0}^\infty e^{-\ell N\mu - \ell^2 \frac{\mu}{2}} \, \frac{(-1)^\ell e^{-\frac{\mu}{2}\ell}}{\prod_{m=1}^\ell (1-x^m)},
\ee
which is a D-brane instanton sum with $A = \mu$ and $\gamma = \frac{\mu}{2}$, up to coefficients with subleading powers in $\ell$.\footnote{As explained in Footnote \ref{foot: instanton coeff}, these subleading powers are allowed and do not alter the natural boundary or the region in which $\Disc_{\theta_2} B_2$ is defined.} We have, in particular, $\re\gamma>0$. The level-2 discontinuity $\Disc_{\theta_2} B_2(\zeta_2)$ is defined only in the exterior \eqref{eq: one-loop exterior} of the natural boundary
\be \label{eq: half-BPS circle}
\left| \zeta_2 + \frac{\mu}{4} \right| = \frac{\mu}{4}
\ee
on the $\zeta_2$-plane.

Since it holds that $\frac{\pa}{\pa N}\log e^{-N^2\zeta_2} = -2 N \zeta_2$, we can extract the leading ``action'' $\zeta_2$ governing the asymptotics of $Z_N$ at large $N$ as
\be \label{eq: extract action}
\zeta_2 = -\frac{1}{2N}\frac{\pa}{\pa N}\log\frac{Z_N}{Z_\infty} + O(N^{-1}).
\ee
For $\re N > 0$, the half-BPS partition function does not admit any non-trivial leading saddle for any fixed $\mu>0$, i.e.\ $Z_N/Z_\infty \to 1$ in the large $N$ limit, and we have $\zeta_2 = 0$ in \eqref{eq: extract action}. Let us now consider instead the analytic continuation of $Z_N$ to $\re N < 0$ via \eqref{eq: half-BPS cont}. While $\re N<0$ is not a physical regime, it is natural from the perspective of Stokes phenomena to consider the asymptotics of $Z_N$ as a function of the phase $\theta = \arg \frac{1}{N}$ of the coupling $g = 1/N$. The singularities on the Borel plane can govern the asympotics of a physical quantity at complex $g$ even when their contributions vanish at positive $g$. We have in this regime
\be
\frac{\pa}{\pa N}\log\frac{Z_N}{Z_\infty} = \sum_{n=1}^\infty \frac{\mu}{e^{\mu(N+n)}-1} = \mu \re N + O(1).
\ee
The action formula \eqref{eq: extract action} tells us that $Z_N$ behaves as $\sim e^{-N^2 \zeta_2}$ where 
\be
\zeta_2 = -\frac{\mu}{2}\frac{\re N}{N} = -\frac{\mu}{4}\left( 1 + e^{2i\theta} \right),
\ee
where $N = |N| e^{-i\theta}$. These values lie precisely on the natural boundary \eqref{eq: half-BPS circle}: the boundary circle is traced once as $\theta$ ranges from $\pi/2$ to $3 \pi/2$. 

All in all, we learn from this example that the dense set of singularities on the natural boundary in $\zeta_2$ represents the asymptotics of $Z_N$ at analytically continued values of $N$. We expect the same interpretation to hold for holographic observables that admit a black hole phase, such as the $\frac{1}{16}$-BPS index and the thermal partition function. While these singularities may not contribute to the asymptotics of $Z_N$ at large positive $N$, they nevertheless obstruct the resummation of $Z_N$ above the black hole threshold. It would be very interesting to understand whether these dense singularities admit a bulk description.

\section{General D-brane instanton data}
\label{sec: general}

\subsection{Summability obstruction at higher loop}
\label{subsec: general obstruction}

In Section \ref{subsec: obstruction}, we found that the level-2 Borel discontinuity $\Disc_{\theta_2} B_2$ is given by a sum over the $\ell$ D-brane instanton sectors and that the sparseness in the coefficients of this sum, via Fabry, results in a natural boundary on the $\zeta_2$-plane. For $\re \gamma<0$, we found that $\Disc_{\theta_2} B_2$ is defined only within a bounded region $\cR$ enclosed by this natural boundary. This property obstructs the summability of large $N$ transseries that contain these D-brane instantons.

Let us now generalize these considerations to the more general situation $F_{0,b \geq 3} \neq 0$. We write the level-2 Borel discontinuity as a sum over the $\ell$ D-brane instanton sector contributions $b_\ell(\zeta_2)$ as 
\be \label{eq: general B2 sum}
\Disc_{\theta_2} B_2(\zeta_2) = \sum_{\ell=1}^\infty b_\ell(\zeta_2).
\ee
We will first determine $b_\ell(\zeta_2)$ individually and then study the properties of their sum $\Disc_{\theta_2} B_2(\zeta_2)$. A stack of $\ell$-coincident D-brane instantons contributes to $Z_N$ as $e^{-N^2 F_0(\ell/N)}$ at genus zero, as explained in Section \ref{subsec: instantons}. We can thus ask that the level-2 Laplace integral of $b_\ell(\zeta_2)$ return this quantity: $N^2 \int_0^{e^{i \theta_2} \infty} e^{-N^2 \zeta_2}\, b_\ell(\zeta_2) \, d\zeta_2 = e^{-N^2 F_0(\ell/N)}$. Inverting, we have
\be
b_\ell(\zeta_2) = \frac{1}{2\pi i} \int_{c-i\infty}^{c+i\infty} \frac{ds}{s} \, e^{s \zeta_2 - s F_0 (\ell/\sqrt{s})},
\ee
where $c>0$ and we set $s = N^2$. The analytic properties of the sum \eqref{eq: general B2 sum} are governed by the large $\ell$ behavior of $b_\ell(\zeta_2)$. To expose the large $\ell$ asymptotics, it is helpful to change the integration variable to $x = \frac{\ell}{\sqrt{s}} = \frac{\ell}{N}$:
\be
b_\ell(\zeta_2) = \frac{1}{\pi i} \int_\cC \frac{dx}{x} \, \exp\! \left[ \ell^2 \, \frac{\zeta_2 - F_0(x)}{x^2} \right],
\ee
where $\cC$ is the image of the previous contour under change of variables. At large $\ell$, this integral is dominated by the saddle point
\be \label{eq: x saddle}
\zeta_2 = F_0(x_*) - \frac{x_*}{2} F_0'(x_*) = \frac{A}{2} x_* - \sum_{b=3}^\infty \frac{b-2}{2} F_{0,b} x_*^b.
\ee
This equation has several solutions $x_*(\zeta_2)$ for $F_{0,b \geq 3} \neq 0$. We take the branch with $x_* \approx \frac{2\zeta_2}{A}$ near $\zeta_2 = 0$ as this is the branch connected to the annulus case. We thus find that $b_\ell(\zeta_2) \sim \mathsf{q}^{\ell^2}$, with
\be \label{eq: general q}
\mathsf{q} = e^{-\mathsf{t}} = \exp\! \left[ - \frac{F_0'(x_*)}{2x_*} \right] = \exp\! \left[ - \left( \frac{A}{2x_*} + \gamma + \sum_{b=3}^\infty \frac{b}{2} \, F_{0,b} \, x_*^{b-2} \right) \right].
\ee
Solving for $x_*$ order by order, we have
\be \label{eq: general q plug}
\mathsf{q} = e^{-\mathsf{t}} = \exp\! \left[ - \left( \frac{A^2}{4\zeta_2} + \gamma + \frac{2F_{0,3}}{A} \, \zeta_2 + \frac{4F_{0,4}}{A^2} \, \zeta_2^2 + \left( \frac{8F_{0,5}}{A^3} + \frac{4F_{0,3}^2}{A^4} \right) \zeta_2^3 + O(\zeta_2^4) \right) \right].
\ee
This agrees with the $\mathsf{q}$ of \eqref{eq: one-loop q} for $F_{0,b \geq 3} = 0$. 

Let us now find the region $\cR$ on the $\zeta_2$-plane in which the sum \eqref{eq: general B2 sum} converges and examine whether its boundary $\pa\cR$ corresponds to a natural boundary of $\Disc_{\theta_2} B_2(\zeta_2)$. Since $b_\ell(\zeta_2) \sim \mathsf{q}^{\ell^2}$ at large $\ell$, the sum converges for $|\mathsf{q}| < 1$, i.e.\ $\re\mathsf{t} > 0$. Using that $F_0'(x_*) = 2\mathsf{t} x_*$ from above, the saddle point equation \eqref{eq: x saddle} becomes
\be \label{eq: general boundary}
\zeta_2 = F_0(x_*) - \mathsf{t} x_*^2.
\ee
The boundary $\pa\cR$ of the convergence region is parametrized by the imaginary values of $\mathsf{t}$ in this saddle equation, with $x_*$ solved accordingly. As we take $\mathsf{t} \to \pm i\infty$, we have $x_* \approx \frac{A}{2\mathsf{t}} \to 0$ and thus $\zeta_2 \to 0$. This point corresponds to the action of thermal AdS $\zeta_2 = I_{\rm AdS} = 0$. At $\mathsf{t} = 0$, we find that $\zeta_2 = F_0(x_*)$ with $F_0'(x_*) = 0$. This point corresponds to the action of the backreacted saddle $\zeta_2 = I = F_0(x_*)$ of Section \ref{subsec: instantons}. Thus, even for $F_{0,b \geq 3} \neq 0$, the boundary $\pa\cR$ of $\cR$ includes points on the $\zeta_2$-plane corresponding to both saddles.

Is $\pa\cR$ a natural boundary? Unlike in the $F_{0,b \geq 3} = 0$ case, the Fabry gap theorem does not apply directly because the sum \eqref{eq: general B2 sum} (or its derivative) is no longer just a power series in $\mathsf{q}$. Explicitly, including fluctuations around the large $\ell$ saddle, we have
\be \label{eq: general b leading}
b_\ell(\zeta_2) \sim \frac{g(\zeta_2)}{\ell} \, \mathsf{q}^{\ell^2},
\ee
with large $\ell$ corrections given by coefficients that depend on both $\ell$ and $\zeta_2$. We can nevertheless show that the large $\ell$ expansion of $b_\ell(\zeta_2)$ has a dense set of singularities on $\pa\cR$. Consider a point on $\pa\cR$ where $\mathsf{q} = e^{2\pi i m/n}$, at which the phases of $\mathsf{q}^{\ell^2}$ are periodic in $\ell$ with period $n$. As we approach it from inside $\cR$, the leading term diverges logarithmically as
\be
\sum_{\ell} \frac{1}{\ell} \mathsf{q}^{\ell^2} \sim -\frac{1}{2} \bigg( \frac{1}{n} \sum_{r=0}^{n-1} e^{2\pi i m r^2/n} \bigg) \log \re\mathsf{t} .
\ee
For coprime $m$ and $n$, the sum over phases vanishes only for $n = 2$ mod $4$. The other roots of unity are still dense on $|\mathsf{q}| = 1$, so these logarithmic singularities are dense on $\pa\cR$. Then as long as our saddle point approximation about $x_*$ remains valid near $\pa\cR$, we have that $\pa\cR$ is a natural boundary of $\Disc_{\theta_2} B_2(\zeta_2)$.

The criterion that made ordinary Borel resummation fail (and thus acceleration necessary) in the case $F_{0,b \geq 3} = 0$ was that the annulus coefficient have a negative real part: $\re\gamma < 0$. A criterion in the case $F_{0,b \geq 3} \neq 0$ that naturally generalizes $\re\gamma<0$ is the following. Since the $\ell$-instanton sector contributes to $Z_N$ as $e^{-N^2 F_0(\ell/N)}$ at genus zero, the D-brane instanton sum diverges and requires acceleration when
\be \label{eq: general criterion}
\re F_0(x) \to -\infty \quad \text{as} \quad x \to +\infty.
\ee
For $F_{0,b \geq 3} = 0$, this condition becomes $\re\gamma < 0$. Accordingly, our proposal in Section \ref{subsec: instantons} becomes here that, in the black hole phase, there exists a family of D-brane instantons satisfying \eqref{eq: general criterion}. In the annulus case, the condition $\re\gamma < 0$ made $\cR$ the bounded disk \eqref{eq: one-loop disk} rather than the unbounded exterior \eqref{eq: one-loop exterior}. This also persists for sufficiently small $F_{0,b \geq 3}$, since $\pa\cR$ in \eqref{eq: general boundary} then stays close to the circle bounding \eqref{eq: one-loop disk}. Beyond this regime, we will assume that $\cR$ remains bounded. Then, as in Section \ref{subsec: obstruction}, the level-2 Laplace integral cannot be performed along any ray going out to infinity.

\subsection{D-brane instanton data on backreacted saddles}
\label{subsec: general dual}

We now relate the D-brane instantons on thermal AdS to those on the backreacted saddle by generalizing the considerations of Section \ref{subsec: backgrounds}. Since the D-brane instanton sectors contribute to the level-2 Borel discontinuity $\Disc_{\theta_2} B_2(\zeta)$ as $b_\ell(\zeta_2) \sim \frac{1}{\ell}g(\zeta_2) \mathsf{q}^{\ell^2}$, we can repeat the steps of Section \ref{subsec: backgrounds} here with $\mathsf{t}$ defined as \eqref{eq: general q}. Poisson summation then turns the sum over $\ell$ with $e^{-\mathsf{t}\ell^2}$ into a sum over $k$ with $e^{-\tilde{\mathsf{t}} k^2}$ with $\tilde{\mathsf{t}} = \frac{\pi^2}{\mathsf{t}}$, in addition to other terms that do not take the form of instanton sectors. We again interpret the sum over $k$ as the D-brane instanton sectors on the backreacted saddle, with its corresponding level-2 singularity at $\tilde\zeta_2 = \zeta_2 - I =0$.

Given $F_0(x)$ of the D-brane instantons on the vacuum saddle, we would like to obtain the genus zero free energy $\tilde F_0(\tilde{x})$ of the D-brane instantons on a backreacted saddle, where we write $x = \ell/N$ and $\tilde x = k/N$. For this, it is helpful to recall the two saddle point relations from the previous subsection
\be
F_0'(x) = 2\mathsf{t} x, \qquad \zeta_2 = F_0(x) - \mathsf{t} x^2.
\ee
We have dropped the asterisks, with the understanding that the above relations hold at the saddle point. The first relation defines $\mathsf{t}$ in \eqref{eq: general q}, and the second relation is the saddle point equation \eqref{eq: x saddle} with $\mathsf{t}$ substituted in. Varying the latter and using the former, we get
\be
d\zeta_2 = (F_0'(x) - 2 \mathsf{t} x ) \, dx - x^2\, d \mathsf{t} = - x^2\, d \mathsf{t}.
\ee
Let us now apply this to the D-brane instantons on the backreacted saddle. The corresponding saddle point $\tilde x$ obeys the same relations with tildes. Using $d\tilde\zeta_2 = d\zeta_2$ and $\tilde{\mathsf{t}} = \frac{\pi^2}{\mathsf{t}}$, we have
\be
\tilde x^2 = -\frac{d\tilde\zeta_2}{d\tilde{\mathsf{t}}} = -\frac{\mathsf{t}^2}{\pi^2} \, x^2.
\ee
We take the root $\tilde x = \frac{1}{i \pi} \mathsf{t} x$, as this choice reproduces \eqref{eq: dual data 2} for $F_{0,b\geq3}=0$. Substituting this into $F_0'(x) = 2\mathsf{t} x$ and its tilded counterpart $\tilde F_0'(\tilde x) = 2\tilde{\mathsf{t}} \tilde x$, we find
\be \label{eq: s dual}
F_0'(x) = 2 \pi i \tilde x, \qquad \tilde F_0'(\tilde x) = -2 \pi i x.
\ee
These relations are symmetric between the two backgrounds. In particular, $\tilde F_0$ is stationary where $x = 0$, i.e.\ at $\tilde x = \frac{F_0'(0)}{2\pi i} = \frac{A}{2\pi i}$. Since we have $\tilde x = \frac{1}{i\pi} \mathsf{t} x$, the saddle point of $\tilde F_0$ corresponds to $\mathsf{t} \to \infty$ (i.e.\ $\tilde{\mathsf{t}} \to 0$) where we have $\zeta_2 \to 0$ as we saw in Section \ref{subsec: general obstruction}. Since $\tilde x$ stays finite as we take $\tilde{\mathsf{t}} \to 0$, the dual saddle point equation $\tilde\zeta_2 = \tilde F_0(\tilde x) - \tilde{\mathsf{t}} \tilde x^2$ gives the stationary value $\tilde F_0 = \tilde\zeta_2 = \zeta_2 - I \to -I$. This shows that a stack of order $N$ D-brane instantons on the backreacted saddle will backreact back into thermal AdS.

Relations of the form \eqref{eq: s dual} are known in matrix models and topological string theory, where they describe an $S$-transformation exchanging a filling fraction with its conjugate \cite{Eynard:2008he,Aganagic:2006wq} (see also \cite{Bonnet:2000dz,Eynard:2008yb,Marino:2008vx}). Here, $x = \ell/N$ plays the role of the filling fraction and $F_0'(x) = 2\pi i \tilde x$ that of its conjugate. In particular, when the backreacted saddle is a black hole, the number $\ell$ of D-brane instantons on thermal AdS is dual to the number $k$ of D-brane instantons on the black hole. In our work, we found these relations on the level-2 Borel plane upon acceleration, so they hold even when the D-brane instanton sum diverges due to the condition \eqref{eq: general criterion} in the large $N$ transseries. The formulas \eqref{eq: s dual} thus relate the genus zero D-brane instanton data on the two backgrounds despite the analytic obstruction to the summability of the large $N$ transseries.

We can find the D-brane instanton data on the backreacted saddle by expanding $\tilde F_0(\tilde x)$ at $\tilde x = 0$, where the relations \eqref{eq: s dual} yield $F_0'(\ell_*/N) = 0$ as in Section \ref{subsec: instantons}. Using $\tilde F_0'(\tilde x) = -2\pi i x$ and $\frac{dx}{d\tilde x} = \frac{2\pi i}{F_0''(x)}$ from \eqref{eq: s dual}, we find
\be \label{eq: general dual data}
\tilde A = -2\pi i \frac{\ell_*}{N}, \qquad \tilde\gamma = \frac{2\pi^2}{F_0''(\ell_*/N)}, \qquad \tilde F_{0,3} = -\frac{4\pi^3 i}{3} \, \frac{F_0'''(\ell_*/N)}{F_0''(\ell_*/N)^3},
\ee
and for $b \geq 2$, we have
\be \label{eq: general dual data b}
\tilde F_{0,b} = -\frac{(2\pi i)^b}{b!} \left[ \left( \frac{1}{F_0''(x)} \frac{d}{dx} \right)^{b-2} \frac{1}{F_0''(x)} \right]_{x = \frac{\ell_*}{N}}.
\ee
For $F_{0,b \geq 3} = 0$, the above expressions reduce to the $(\tilde A, \tilde\gamma)$ of \eqref{eq: dual data 2}. These formulas relate the D-brane instanton data on thermal AdS to that on the backreacted saddle.

More explicitly, let us expand to first order in the $F_{0,b \geq 3}$ at fixed $A$ and $\gamma$. Solving $F_0'(\ell_*/N) = 0$ to this order, we have
\be
\frac{\ell_*}{N} = -\frac{A}{2\gamma} - \frac{1}{2\gamma} \sum_{b=3}^\infty b \, F_{0,b} \left( -\frac{A}{2\gamma} \right)^{b-1} + O(F_{0,b \geq 3}^2).
\ee
This determines $\tilde A$ via \eqref{eq: general dual data}. The backreacted action $I = F_0(\ell_*/N)$ is
\be \label{eq: general I}
I = -\frac{A^2}{4\gamma} + \sum_{b=3}^\infty F_{0,b} \left( -\frac{A}{2\gamma} \right)^b + O(F_{0,b \geq 3}^2).
\ee
Evaluating also the higher terms, we find
\ie \label{eq: general dual data expanded}
\tilde\gamma &= \frac{\pi^2}{\gamma} - \frac{\pi^2}{\gamma^2} \sum_{b=3}^\infty \frac{b(b-1)}{2} \, F_{0,b} \left( -\frac{A}{2\gamma} \right)^{b-2} + O(F_{0,b \geq 3}^2) \\
\tilde F_{0,c} &= \left( \frac{i\pi}{\gamma} \right)^c \sum_{b=c}^\infty \frac{b!}{c!\,(b-c)!} \, F_{0,b} \left( -\frac{A}{2\gamma} \right)^{b-c} + O(F_{0,b \geq 3}^2)
\fe
for $c \geq 3$.

\subsection{Precision bound beyond the annulus}
\label{subsec: beyond}

With region $\cR$ as defined in Section \ref{subsec: general obstruction}, the proposal of Section \ref{subsec: obstruction} applies as stated in \eqref{eq: geometric Shat}--\eqref{eq: precision bound norm}. Let us ask how the higher open string corrections $F_{0,b\geq3}$ can affect the precision scale $\hat I$.

The precision scale $\hat I$ is set by the real part of $\zeta_2$ at the rightmost point of $\cR$. To find this point, consider moving along the boundary $\pa\cR$ where $\mathsf{t}$ is pure imaginary, as in the discussion around \eqref{eq: general boundary}. We can ask when $\re\zeta_2$ becomes stationary. Since $d\zeta_2 = -x^2\, d\mathsf{t}$ with $d\mathsf{t}$ pure imaginary, we find that $\re\zeta_2$ is stationary when $x^2$ is real. We will assume that the rightmost point has real $x$ as in the annulus case.\footnote{In the annulus case, $x = \frac{\re A}{2|\re\gamma|}$ which gives \eqref{eq: one-loop Shat} when plugged into $\re\zeta_2 = \re F_0(x)$.} Then, we have that $\mathsf{t}x^2$ and $2\mathsf{t}x$ are pure imaginary, so the saddle point relations $\zeta_2 = F_0(x) - \mathsf{t} x^2$ and $F_0'(x) = 2 \mathsf{t} x$ give
\be
\re\zeta_2 = \re F_0(x), \qquad \re F_0'(x) = 0.
\ee
The rightmost point of $\cR$ thus corresponds to a stationary point of $\re F_0(x)$ along the real $x$-axis, and $\hat I$ is the value of $\re F_0$ at that point. Let us suppose that $\re A \geq 0$ and that $\re F_0(x)$ has a single maximum on $x \geq 0$, beyond which it decreases without bound, as happens in the annulus case. Then we have
\be \label{eq: general Shat}
\hat I = \max_{x \geq 0} \re F_0(x).
\ee
Since the $\ell$-instanton sector on thermal AdS has magnitude $e^{-N^2 \re F_0(\ell/N)}$, the precision bound $e^{-N^2\hat I}$ is the size of the smallest term of the instanton sum. On the backreacted saddle, we can apply the same argument. At the rightmost point, $\tilde{\mathsf{t}} = \frac{\pi^2}{\mathsf{t}}$ is pure imaginary and $\tilde x = \frac{1}{i\pi}\mathsf{t} x$ is real, so we get $\hat I - \re I = \re\tilde F_0(\tilde x)$. Under the same assumptions on $\tilde F_0$, the relative precision bound $e^{-N^2(\hat I - \re I)}$ is then the size of the smallest term of the instanton sum on the backreacted saddle. As in Section \ref{subsec: obstruction}, the optimal truncation scales of the instanton sums around the two saddles thus agree with the precision scale.

For real $x$, we have that $\re F_0(x)$ is $F_0(x)$ with all its coefficients replaced by their real parts. The maximum \eqref{eq: general Shat} is therefore the backreacted action $I$, i.e.\ the stationary value of $F_0$, with the same real replacement. To first order in the $F_{0,b \geq 3}$, \eqref{eq: general I} gives
\be \label{eq: general Shat expanded}
\hat I = \frac{(\re A)^2}{4|\re\gamma|} + \sum_{b=3}^\infty \re F_{0,b} \left( \frac{\re A}{2|\re\gamma|} \right)^b + O(F_{0,b \geq 3}^2).
\ee
The first term is the annulus result \eqref{eq: one-loop Shat}. The sum provides the leading set of open string corrections. Note that $\hat I = 0$ when we have instantons on the AdS saddle with pure imaginary action: $\re A = 0$. This statement persists also with higher power corrections in $F_{0,b \geq 3}$. In that case, the precision bound becomes of order $1$ on the $e^{-N^2}$ scale. We will return to this point shortly.

\section{Expectations for generic situations}
\label{sec: generic}

Consider the thermal partition function $Z_N$ of a theory of $N \times N$ matrices in the black hole phase, in a situation that is sufficiently generic. What precision bound should we expect for its gravitational transseries $Z_N^{\rm grav}$?

Let us suppose that, in generic situations, $Z_N^{\rm grav}$ admits contributions from different families of D-brane instantons, labeled by $i$. (This is the case even in simple large $N$ matrix models, where one gets a family of multi-eigenvalue instantons for each non-trivial saddle of the effective potential \cite{Marino:2012zq}.) As we saw earlier, each such family $i$ can obstruct resummation with its own precision scale set by $\hat I_i \geq 0$. The family producing the largest remainder then sets the precision bound at the scale $e^{-N^2 \hat I}$, with
\be \label{eq: generic Shat}
\hat I = \min_i \hat I_i.
\ee
It follows that a single family with $\hat I_i = 0$ is enough to set $\hat I = 0$, which is the coarsest precision allowed by our proposal.

When do we get $\hat I = 0$? According to \eqref{eq: general Shat}, a family has $\hat I_i = 0$ when $\re F_0(\ell/N) \leq 0$ for all $\ell$, i.e.\ when no term of its instanton sum on thermal AdS is smaller than thermal AdS. This requires a pure imaginary D-brane instanton action $\re A_i = 0$ on thermal AdS, and a single such family in the black hole phase would be sufficient to have $\hat I = 0$. We can state the analogous condition on the black hole. By \eqref{eq: s dual}, $\ell = 0$ corresponds to the stationary point $k = \frac{NA}{2\pi i}$ of the genus zero free energy $\tilde F_0(k/N)$ on the black hole, i.e.\ to the stack that backreacts into thermal AdS. Thus, if $A$ is pure imaginary then $k$ is real and vice versa, which would lead to $\hat I_i = 0$. A real $k$ can arise, for example, if $\tilde F_0$ is real with $\tilde A > 0$ and if the instanton sum diverges as in a tilded version of \eqref{eq: general criterion}. If so, $\tilde F_0(k/N)$ will increase from $0$ at $k = 0$ to a maximum at a real $k > 0$ and then decreases without bound at large $k$. This maximum corresponds to the stationary point $k = \frac{NA}{2\pi i}$ above, so we get $\hat I_i = 0$. In the $F_{0,b\geq3} = 0$ case, these conditions become $\tilde A > 0$ and $\tilde\gamma < 0$.

With such D-brane instantons, the precision bound of Section \ref{subsec: obstruction} becomes
\be \label{eq: generic bound}
|Z_N - \overline{Z_N^{\rm grav}}| \, \gtrsim \, O(1), \qquad \frac{|Z_N - \overline{Z_N^{\rm grav}}|}{|Z_N|} \, \gtrsim \, O\big( e^{-N^2 |\re I_{\rm BH}|} \big)
\ee
on the $e^{-N^2}$ scale. In the black hole phase, the gravitational transseries $Z_N^{\rm grav}$ determines $Z_N$ only up to the size of the thermal AdS contribution.

We can compare this result with what can be expected from the double-cone wormhole \cite{Saad:2018bqo}. Since its classical action vanishes, the double-cone wormhole contributes to the two-boundary amplitude an order $1$ quantity on the $e^{-N^2}$ scale. Suppose that this contribution computes the variance $\la |\delta_N|^2 \ra_N$ of the erratic part $\delta_N$ in the decomposition $Z_N \sim Z_N^{\rm grav} + \delta_N$. Then $|\delta_N|$ would be of order $1$. Since $\delta_N$ is not captured by $Z_N^{\rm grav}$ in the above decomposition, the gravitational transseries could determine $Z_N$ at best up to an error of order $1$. This agrees with the bound \eqref{eq: generic bound} set by a family of D-brane instantons for which $\hat I_i = 0$. Whether such D-brane instantons are generic above the black hole threshold is an open question.

\section{Discussion}

We conclude with a partial list of open questions:
\begin{enumerate}

\item Is $\hat I = 0$ generic?

\item What are the implications of the proposed precision bound for the microcanonical spectrum at energies of order $N^2$?

\item The precision bound states that the gravitational transseries $Z_N^{\rm grav}$ determines $Z_N$ only up to a remainder of order $e^{-N^2\hat I}$, but it does not necessarily say that this remainder is erratic in $N$. Is the remainder erratic in $N$ in sufficiently generic situations? If so, does the shape of the natural boundary determine its statistics?

\item In Section \ref{subsec: general dual}, we found that stacks of order $N$ D-brane instantons take thermal AdS to a saddle $M^*$ and $M^*$ back to thermal AdS. But the derivation of \eqref{eq: s dual} does not assume that the starting saddle is thermal AdS. Can other families of D-brane instantons relate other pairs of saddles in the same way, such as between two black holes?

\item The critical acceleration level \eqref{eq: critical level}
\be
\frac{1}{k'} = \frac{1}{k} \left( 1 - \frac{1}{p} \right)
\ee
has an interesting implication in M-theoretic examples of holography. Set $k=1$. Consider the growth exponent $p=\frac{3}{2}$, as would arise from having $\ell$ M2 instantons with worldvolume factors $e^{-\ell^{3/2}}$. This gives the acceleration level $k'=3$ consistent with the leading large $N$ behavior $e^{-N^3}$ of the worldvolume theory of $N$ M5 branes. Conversely, we can consider the growth exponent $p=3$ as would be expected for $\ell$ M5 instantons with worldvolume factors $e^{-\ell^{3}}$. This gives $k'=\frac{3}{2}$ consistent with the leading large $N$ behavior $e^{-N^{3/2}}$ of the worldvolume theory of $N$ M2 branes.

\item What is the physical origin of the dense singularities forming the natural boundary on the $\zeta_2$-plane, along the lines of Section \ref{subsec: dense}?

\item It would be of importance to extend the applications of higher Borel-\'{E}calle resummation, i.e.\ accelero-summation, to more examples and to study the extent of its physical implications.

\item The symmetric structure of singularities on the Borel plane of the large $N$ expansion indicates the presence of ``negative'' brane instantons of weight $e^{+1/g_s}$ \cite{Okuda:2006fb,Aniceto:2011nu,Dijkgraaf:2016lym,Marino:2022rpz,Schiappa:2023ned}. It would be interesting to study their consequences on the higher Borel plane.

\item What could be the bulk origin of the remainder term, and is it reasonable to hope for a bulk formalism that captures it?

\end{enumerate}

\section*{Acknowledgments} 

I would like to thank Jan Boruch, Yiming Chen, Davide Gaiotto, Jaume Gomis, Luca Iliesiu, Jonah Kudler-Flam, Juan Maldacena, Chintan Patel, Eric Perlmutter, Ricardo Schiappa, Arvin Shahbazi-Moghaddam, Stephen Shenker, Douglas Stanford, Edward Witten, Zhenbin Yang, and Yanjun Zhou for valuable discussions, and especially Yiming Chen, Ricardo Schiappa, Douglas Stanford, and Zhenbin Yang for particularly insightful conversations. I am grateful to the organizers of Quantum Information in Quantum Gravity 2026 (QIQG 2026), physics group meeting at the Institute for Advanced Study, and LITP seminar at Stanford University for the opportunities to present versions of this work.

The work of the group at ETH is supported by the Simons Foundation grant 994306 (Simons Collaboration on Confinement and QCD Strings) and by the NCCR SwissMAP funded by the Swiss National Science Foundation. The author used ChatGPT and Claude to assist with improving mathematical rigor in various parts and manuscript revisions. The scientific content, arguments, calculations, and conclusions presented in this work are the author’s own.

\bibliography{references}

\bibliographystyle{utphys}

\end{document}